\documentclass[trackchanges,twocolumn, twocolappendix]{aastex701}

\usepackage{amsmath}
\usepackage{booktabs}
\usepackage{enumitem}

\begin{document}

\title{Constructing a Mock Galaxy Catalog for the All-sky SPECtroscopic Survey of Nearby Galaxies (A-SPEC) Using the Machine-assisted Semi-Simulation Model}

\author[0000-0003-4127-6110]{Dongkok Kim}
\affiliation{Astronomy Program, Department of Physics and Astronomy, Seoul National University, 1 Gwanak-ro, Gwanak-gu, Seoul 08826, Republic of Korea}
\affiliation{Institute for Data Innovation in Science, Seoul National University, Seoul 08826, Korea}
\email[show]{cruithne33@snu.ac.kr}  

\author[0000-0003-3977-1761]{Yongseok Jo} 
\affiliation{NSF-Simons AI Institute for the Sky (SkAI), 172 E. Chestnut St., Chicago, IL 60611, USA}
\email{me@yongseokjo.com}  

\author[0000-0003-3428-7612]{Ho Seong Hwang}
\affiliation{Astronomy Program, Department of Physics and Astronomy, Seoul National University, 1 Gwanak-ro, Gwanak-gu, Seoul 08826, Republic of Korea}
\affiliation{SNU Astronomy Research Center, Seoul National University, 1 Gwanak-ro, Gwanak-gu, Seoul 08826, Republic of Korea}
\affiliation{Korea Astronomy and Space Science Institute, 776 Daedeok-daero, Yuseong-gu, Daejeon 34055, Republic of Korea}
\email[show]{hhwang@astro.snu.ac.kr}  

\author[0000-0003-4464-1160]{Ji-hoon Kim}
\affiliation{Center for Theoretical Physics, Department of Physics and Astronomy, Seoul National University, Seoul 08826, Republic of Korea}
\affiliation{Institute for Data Innovation in Science, Seoul National University, Seoul 08826, Korea}
\affiliation{SNU Astronomy Research Center, Seoul National University, 1 Gwanak-ro, Gwanak-gu, Seoul 08826, Republic of Korea}
\email{mornkr@snu.ac.kr}  

\author[0000-0002-4391-2275]{Juhan Kim}
\affiliation{Center for Advanced Computation, Korea Institute for Advanced Study, 85 Heogiro, Dongdaemun-gu, Seoul, 02455, Republic of Korea}
\email{kjhan0606@gmail.com}

\author[0000-0002-6810-1778]{Jaehyun Lee}
\affiliation{Korea Astronomy and Space Science Institute, 776 Daedeok-daero, Yuseong-gu, Daejeon 34055, Republic of Korea}
\email{syncphy@gmail.com}  

\author[0009-0002-9878-1126]{Hyeonguk Bahk}
\affiliation{Astronomy Program, Department of Physics and Astronomy, Seoul National University, 1 Gwanak-ro, Gwanak-gu, Seoul 08826, Republic of Korea}
\email{spica095@snu.ac.kr}  

\author{Young-Man Choi}
\affiliation{Department of Mechanical Engineering, Ajou University, 206 Worldcup-ro, Yeongtong-gu, Suwon-si, Gyeonggi-do, 16499, Republic of Korea}
\email{ymanchoi@ajou.ac.kr}

\author{Moo-Young Chun}
\affiliation{Korea Astronomy and Space Science Institute, 776 Daedeok-daero, Yuseong-gu, Daejeon 34055, Republic of Korea}
\email{mychun@kasi.re.kr}

\author[0000-0002-6154-7558]{Sang-Hyun Chun}
\affiliation{Korea Astronomy and Space Science Institute, 776 Daedeok-daero, Yuseong-gu, Daejeon 34055, Republic of Korea}
\email{shyunc.m@gmail.com}

\author[0000-0002-3043-2555]{Haeun Chung}
\affiliation{University of Arizona, Steward Observatory, 933 N. Cherry Ave., Tucson, AZ 85721, USA}
\email{haeunchung@email.arizona.edu} 

\author[0009-0001-3899-7198]{Kim Dachan}
\affiliation{Department of Astronomy, Space Science and Geology, Chungnam National University, 99 Daehak-ro, Yuseong-gu, Daejeon 34134, Republic of Korea}
\email{zmflvlxm@gmail.com}

\author[0000-0003-4923-8485]{Sungwook E. Hong}
\affiliation{Korea Astronomy and Space Science Institute, 776 Daedeok-daero, Yuseong-gu, Daejeon 34055, Republic of Korea}
\affiliation{Astronomy Campus, University of Science and Technology, 776 Daedeok-daero, Yuseong-gu, Daejeon 34055, Republic of Korea}
\email{swhong@kasi.re.kr}

\author[0000-0003-4738-4251]{Minhee Hyun}
\affiliation{Kavli Institute for Particle Astrophysics and Cosmology, Stanford University, Menlo Park, CA 94305, USA}
\email{minhee20400@gmail.com}

\author[0000-0002-8434-979X]{Donghui Jeong}
\affiliation{Department of Astronomy and Astrophysics, The Pennsylvania State University, University Park, PA 16802, USA}
\affiliation{Institute for Gravitation and the Cosmos, The Pennsylvania State University, University Park, PA 16802, USA}
\affiliation{School of Physics, Korea Institute for Advanced Study, 85 Heogiro, Dongdaemun-gu, Seoul, 02455, Republic of Korea}
\email{djeong@psu.edu}  

\author[0000-0002-1710-4442]{Jae-Woo Kim}
\affiliation{Korea Astronomy and Space Science Institute, 776 Daedeok-daero, Yuseong-gu, Daejeon 34055, Republic of Korea}
\email{kjw0704@kasi.re.kr}  

\author{Kang-Min Kim}
\affiliation{Korea Astronomy and Space Science Institute, 776 Daedeok-daero, Yuseong-gu, Daejeon 34055, Republic of Korea}
\email{kmkim@kasi.re.kr}

\author[0000-0003-0009-5161]{Yunjong Kim}
\affiliation{Korea Astronomy and Space Science Institute, 776 Daedeok-daero, Yuseong-gu, Daejeon 34055, Republic of Korea}
\affiliation{Astronomy Campus, University of Science and Technology, 776 Daedeok-daero, Yuseong-gu, Daejeon 34055, Republic of Korea}
\email{yjkim@kasi.re.kr}

\author[0000-0002-9434-5936]{Jongwan Ko}
\affiliation{Korea Astronomy and Space Science Institute, 776 Daedeok-daero, Yuseong-gu, Daejeon 34055, Republic of Korea}
\email{jwko@kasi.re.kr}

\author[0009-0007-9451-5337]{Minseong Kwon}
\affiliation{Astronomy Program, Department of Physics and Astronomy, Seoul National University, 1 Gwanak-ro, Gwanak-gu, Seoul 08826, Republic of Korea}
\email{mhee7173@snu.ac.kr}  

\author[0000-0002-3808-7143]{Ho-Gyu Lee}
\affiliation{Korea Astronomy and Space Science Institute, 776 Daedeok-daero, Yuseong-gu, Daejeon 34055, Republic of Korea}
\email{hglee@kasi.re.kr}

\author[0000-0003-0283-8352]{Jong Chul Lee}
\affiliation{Korea Astronomy and Space Science Institute, 776 Daedeok-daero, Yuseong-gu, Daejeon 34055, Republic of Korea}
\email{jclee@kasi.re.kr}  

\author[0000-0001-7594-8072]{Yongseok Lee}
\affiliation{Korea Astronomy and Space Science Institute, 776 Daedeok-daero, Yuseong-gu, Daejeon 34055, Republic of Korea}
\email{yslee@kasi.re.kr}  

\author{Hyunho Lim}
\affiliation{Department of Mechanical Engineering, Ajou University, 206 Worldcup-ro, Yeongtong-gu, Suwon-si, Gyeonggi-do, 16499, Republic of Korea}
\email{qwecsz@ajou.ac.kr}

\author[0000-0002-0418-5335]{Heeyoung Oh}
\affiliation{Korea Astronomy and Space Science Institute, 776 Daedeok-daero, Yuseong-gu, Daejeon 34055, Republic of Korea}
\email{hyoh@kasi.re.kr}

\author[0000-0001-9521-6397]{Changbom Park}
\affiliation{School of Physics, Korea Institute for Advanced Study, 85 Heogiro, Dongdaemun-gu, Seoul, 02455, Republic of Korea}
\email{cbp@kias.re.kr}  

\author[0000-0002-4362-4070]{Hyunmi Song}
\affiliation{Department of Astronomy and Space Science, Chungnam National University, Daejeon 34134, Republic of Korea}
\email{yesuane@gmail.com}

\author[0009-0005-9012-6006]{Mingyeong Yang}
\affiliation{Korea Astronomy and Space Science Institute, 776 Daedeok-daero, Yuseong-gu, Daejeon 34055, Republic of Korea}
\affiliation{Astronomy Campus, University of Science and Technology, 776 Daedeok-daero, Yuseong-gu, Daejeon 34055, Republic of Korea}
\email{mmingyeong@kasi.re.kr}

\author[0000-0003-0134-8968]{Yongmin Yoon}
\affiliation{Department of Astronomy and Atmospheric Sciences, College of Natural Sciences, Kyungpook National University, Daegu 41566, Republic of Korea}
\email{yymx2aa@gmail.com}

\begin{abstract}
We present a methodology for constructing a mock galaxy catalog for the All-sky SPECtroscopic survey of nearby galaxies (A-SPEC) using the Machine-assisted Semi-Simulation Model. The model is trained on the cosmological magnetohydrodynamical simulation IllustrisTNG to predict baryonic properties of subhalos from dark-matter-only features and is applied to our own $N$-body simulation tailored to satisfy the requirements of A-SPEC. We have improved the model's accuracy by introducing additional features such as subhalo anisotropy parameters and modified definitions of the subhalo environment, which result in the coefficient of determination $R^2=0.96, 0.90, 0.70, 0.79$ for stellar mass, gas mass, star formation rate, and gas metallicity, respectively. The resulting mock galaxies reproduce the luminosity-dependent clustering of the target galaxies when tuned to match the number density. We discuss avenues for further improvement, including the role of environment in the predictions. We release the mock galaxy catalog with the baryonic properties predicted from the model.

\end{abstract}



\section{Introduction \label{sec:Sec1}}
\subsection{A-SPEC and K-SPEC \label{sec:Sec1.1}}
Observations have shown that the spatial distribution of galaxies in the Universe is nonrandom, characterized by a distinctive pattern of clustering \citep{Hauser1973, Einasto1980, CfA1982, CfA1986}. This galaxy distribution has served as one of the key diagnostics of cosmic evolution in the late Universe \citep{GH1989, Percival2001, Tegmark2006, Abott2018}. Together with observations of the cosmic microwave background \citep{Komatsu2011, Planck2020} and supernovae \citep{Riess1998, Perlmutter1999}, it has played a central role in establishing the current standard cosmological framework, which incorporates dark matter (DM) as the source of gravitational potential and dark energy as the driver of accelerated cosmic expansion. Within this context, numerous galaxy surveys at various wavelengths have been carried out or planned that target different galaxy populations in various cosmic epochs \citep{York2000, Colless2001, Jones2004, LSST2009, Driver2011, Euclid2011, Dore2014, DES2016, DESI2016, Aihara2018}. In particular, measurements of galaxy spectra, which encode intrinsic properties as well as the Hubble flow and peculiar motions, have enabled precise estimates of galaxy distances and physical properties.

Although the majority of the current surveys aim to extend the survey area to higher redshifts, the mapping of the low-redshift Universe has not been complete (see Figure \ref{fig:Fig1_SpecCompl}). The importance of mapping the local Universe within hundreds of megaparsecs lies in providing an unbiased starting point for our understanding of the cosmos. For example, the local galaxy catalog will serve as a pivotal sample for studying dynamical dark energy \citep[e.g.,][]{Dong2023}, the cosmic matter density, and the Hubble tension \citep{Whitbourn2014, Bohringer2020}.
For this purpose, the All-sky SPECtroscopic survey of nearby galaxies (A-SPEC) was designed to compile the physical properties and redshifts of galaxies brighter than $K_S=13.75$ \citep{Kwon2026} and is currently awaiting the commissioning of K-SPEC, the multiobject spectrograph for the survey.

\subsection{Mock Galaxy Catalog \label{sec:Sec1.2}}
Mock galaxy catalogs have played a crucial role in validating survey designs for a given science goal, including the selection function, passband, brightness limit, and survey area \citep[e.g.,][]{Angulo2008, Hamaus2022}. However, owing to cosmic variance and our limited knowledge of astrophysical processes, a full description of the underlying Universe cannot be obtained from observations alone. Here, simulations play a complementary role in bridging this gap. Mock catalogs generated under different cosmological models and/or astrophysical prescriptions enable a direct interpretation of the corresponding observed quantities. Alternatively, a statistical approach can be adopted, in which simulations spanning a range of cosmological models are used to derive the corresponding likelihood functions, which are then compared with observations. In short, mock galaxy catalogs serve not only as a direct reference for comparison with observations but also as a statistical test bed, providing covariances between cosmological parameters that cannot be directly obtained from observations.

Mock galaxy catalogs take various forms, depending on the scientific goal. For example, rapidly generating a large number of approximate catalogs may suffice for estimating parameter covariances \citep{Cole2005, Hutsi2006, Kitaura2016, Forero2025}. Because such catalogs are designed to reproduce statistical measures, however, they typically lack information on individual galaxy properties. By contrast, a single reference mock catalog cannot capture cosmic variance but is well suited for reproducing diverse baryonic properties and precise galaxy clustering \citep{Kim2011, Rodriguez2016, Yuan2022, To2024}.
In this paper, we construct and release a reference mock galaxy catalog with a comprehensive set of baryonic properties, designed to match the one-point and two-point statistics of the A-SPEC target galaxies.

\subsection{Modeling the Galaxy--Halo Connection \label{sec:Sec1.3}}
Under the concordance cosmological framework, density fluctuations in the early Universe evolved through gravity and cosmic expansion, shaping the large-scale structures in the late-time Universe. As the cosmic structure forms and grows, DM collapses into bound objects called halos. Baryons fall into their gravitational potential wells and form galaxies. As galaxies can be directly observed through electromagnetic radiation, exploring their formation and evolution is essential for generating realistic mock galaxies, and hence for cosmological interpretation. Galaxies residing in the central parts of DM halos are primarily characterized by the physical properties of halos such as mass and internal structure, while the surrounding environment and evolution history play a supplementary role. However, the scaling relations between them exhibit large scatter due to degeneracies in the complicated astrophysical effects and diverse formation and evolution histories.

Strategies for modeling the galaxy--halo connection fall into two types, depending on how they include baryonic physics. On the most complex and physics-driven side, we have hydrodynamical simulations \citep{Magneticum2014, EAGLE2015, Springel2018, SIMBA2019, HR52021}. They implement baryons---such as stars, gas, and black holes---as particles and/or cells, and evolve them by solving the coupled equations of gravity and hydrodynamics. Subresolution processes such as star formation, stellar and active galactic nuclei (AGN) feedback, and black hole accretion are incorporated through subgrid prescriptions, together shaping the baryonic field in the universe. Arguably, they have an advantage of directly handling the dynamics of baryons, but they require enormous computational resources to achieve a high resolution. Thus, simulations covering our survey volume (${\sim}2 \,\rm{Gpc}^3$) are limited and suffer from a fine-tuning problem, just as in the other methods described below. An alternative approach that includes baryonic physics while enabling galaxy formation across large volumes is the semianalytical model (SAM). SAMs describe the evolution of baryons with a suite of approximate prescriptions \citep{Cole2000, Croton2006, Lee2013, Croton2016, Cora2018}.

On the other hand, there exist methods that populate the DM field with galaxies without invoking baryonic physics. On the data-driven side, the halo occupation distribution (HOD) defines a probability function for galaxy counts in halos \citep{Zheng2005, Hearin2016} to match observations. Empirical modeling \citep{Moster2018, Behroozi2019, Wechsler2022} and abundance matching \citep{Reddick2013, Lehmann2017} have also been widely adopted to study the galaxy--halo connection and to construct mock data \citep[see][and references therein for a review]{Wechsler2018}.

\begin{figure}[t]
\centering
\includegraphics[width=0.99\linewidth]{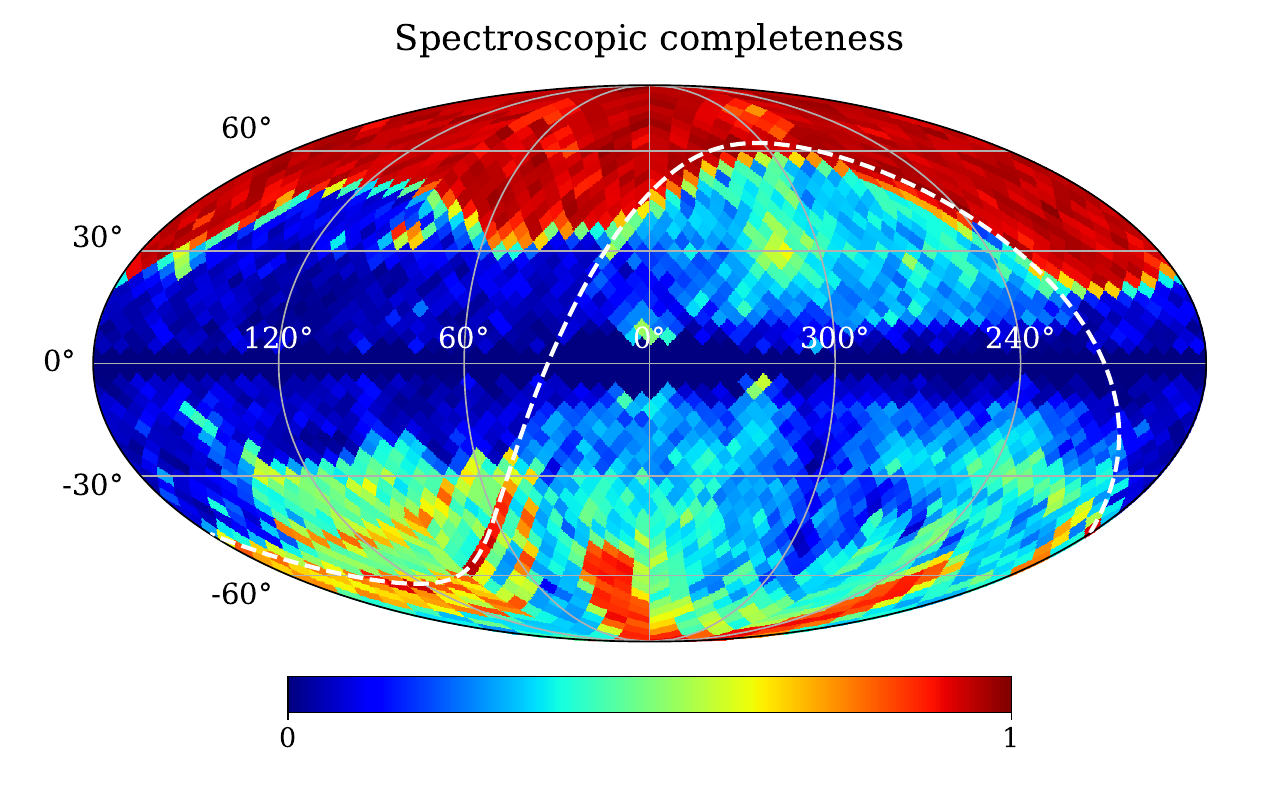}
\caption{Spectroscopic completeness of galaxies with $K_S \leqslant 13.75$ in the Two Micron All Sky Survey extended source catalog in the Galactic coordinate system. The white dashed line indicates the celestial equator.\label{fig:Fig1_SpecCompl}}
\end{figure}

One recent approach to studying the galaxy--halo connection is based on machine learning (ML). A pioneering work by \citet{Kamdar2016a, Kamdar2016b} tested a number of ML algorithms, trained on an SAM and hydrodynamical simulation, to predict the baryonic properties of central galaxies from their DM halo properties. In subsequent studies, additional input features, such as halo environment indicators and DM properties measured at different redshifts together with merger tree information, were employed. They also tested different ML strategies, each aimed at further boosting model accuracy \citep{Agarwal2018, JK2019, Xu2021, Delgado2022, Lovell2022, McGibbon2022, dS2022, Stiskalek2022, Jespersen2022, Hernandez2023, dA2023, Xu2024}.

In this study, we adopt the Machine-assisted Semi-Simulation Model \citep[MSSM;][hereafter JK19]{JK2019} to assign baryonic properties to subhalos using only DM features. The outcome of this halo-painting process serves as the basis for constructing a mock galaxy catalog that mimics the statistical properties of the target galaxies of A-SPEC from an $N$-body simulation. To generate a DM subhalo catalog to which the ML model is applied, we conduct an $N$-body simulation that satisfies the requirements of A-SPEC such as volume and resolution.

The remainder of the paper is organized as follows. In Section \ref{sec:Sec2}, we introduce the target galaxy catalog and the hydrodynamical simulation used to train our machine. Section \ref{sec:Sec3} details the ML strategy adopted in this study, along with the $N$-body simulation to which the ML model is applied. In Section \ref{sec:Sec4}, we present the training and application results of MSSM and compare the resulting galaxy catalog with observations. We discuss future prospects and role of input features in Section \ref{sec:Sec5}. We conclude in Section \ref{sec:Sec6}.\\

\section{Data \label{sec:Sec2}}
\subsection{Target Galaxy Catalog \label{sec:Sec2.1}}
A-SPEC aims to construct a highly complete redshift catalog of galaxies brighter than $K_S=13.75$ (Vega mag) over the entire sky. Thus, we build the target galaxy catalog based on the publicly available Two Micron All Sky Survey (2MASS) extended source catalog \citep[XSC;][]{Jarrett2000}. The 2MASS XSC contains photometry of approximately one million galaxies in near-infrared bands ($J,H,K_S$) and is known to be highly complete up to $K_S=14$ at high Galactic latitudes \citep{Afshordi2004, Rahman2016}. Our galaxy sample is selected using isophotal photometry $K_{20}$, defined at $20\,\rm{mag}~\rm{arcsec}^{-2}$ within an elliptical aperture (k\textunderscore m\textunderscore k20fe). We select galaxies with $K_{20} \leqslant 13.75$ after applying the Galactic extinction correction \citep{SFD1998}.

\begin{figure}[t]
\centering
\includegraphics[width=0.99\linewidth]{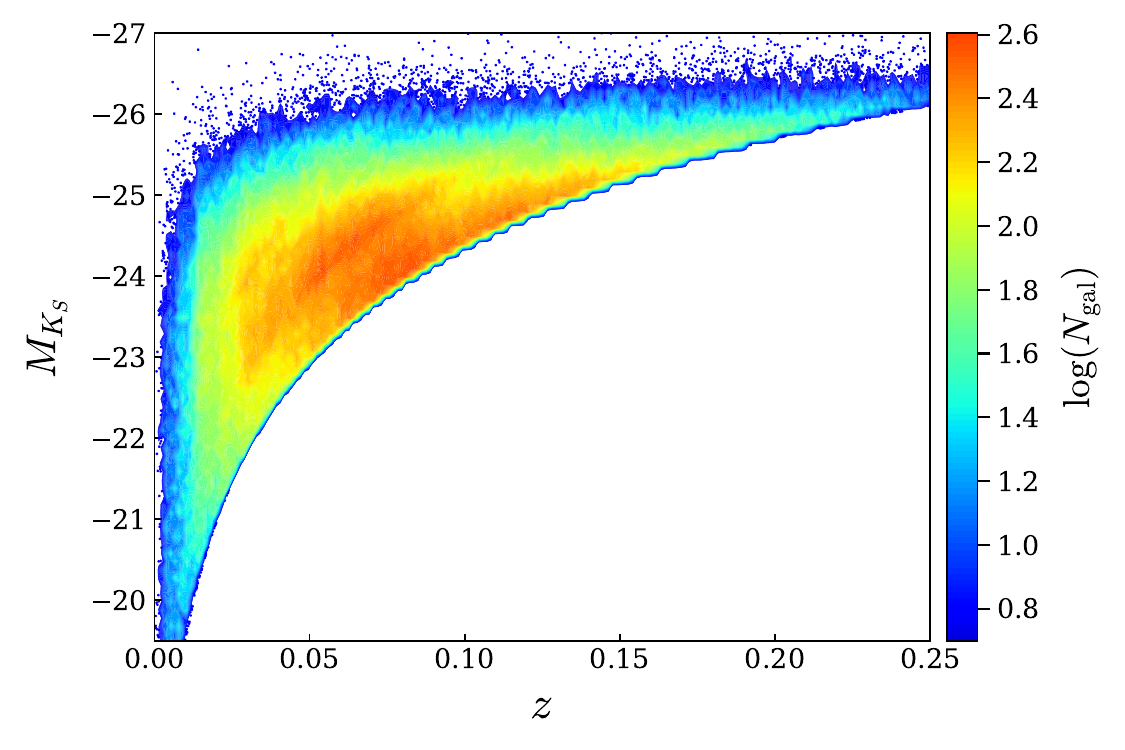}
\caption{Absolute magnitude $M_{K_S}$ of galaxies with $K_S \leqslant 13.75$ as a function of spectroscopic redshift. The color map shows the number of galaxies ($N_{\rm{gal}}$) with measured redshifts in the literature. Signatures from the flux limits of various surveys can be observed as discontinuous strips. The limiting absolute magnitude decreases rapidly, covering over three magnitudes. \label{fig:Fig2_z_MKs}}
\end{figure}

To supplement the target galaxy catalog with redshifts and exclude galaxies without measurements, we compiled redshifts available in the literature, including \citet{Quintana1995}, Updated Zwicky Catalog \citep{Falco1999}, PSCz \citep{Saunders2000}, 2dFGRS \citep{Colless2001}, 6dFGRS \citep{Jones2004}, 2MRS \citep{Huchra2012}, and Sloan Digital Sky Survey DR17 \citep{SDSS2022} supplemented by \citet{Hwang2010}. Figure \ref{fig:Fig1_SpecCompl} shows the spectroscopic completeness of galaxies with $K_S \leqslant 13.75$ in the Galactic coordinate system. Spectroscopic completeness $C$ is computed at $N_{\rm{side}}=16$ HEALPix \citep{Gorski2005} pixels; completeness is defined as the ratio of galaxies that already have redshift measurements. The northern hemisphere exhibits higher completeness, with a mean of $C\approx0.53$, whereas in the southern hemisphere, the mean completeness is $C\approx0.35$ at $|b|>10^{\circ}$\footnote{$b$ denotes the Galactic latitude.}, leaving about 450,000 galaxies as targets for A-SPEC. In Figure \ref{fig:Fig2_z_MKs}, we show the distribution of galaxies with $K_S \leqslant 13.75$ in redshift-absolute magnitude ($M_{K_S}$) space.\footnote{Throughout the paper, we apply K and evolutionary (k+e) correction according to \citet{Bell2003} when converting apparent magnitude to absolute magnitude, or vice versa.} Spectroscopic incompleteness is shown as discontinuous strips representing the flux limits of multiple spectroscopic surveys. 

\begin{figure}[t]
\centering
\includegraphics[width=0.99\linewidth]{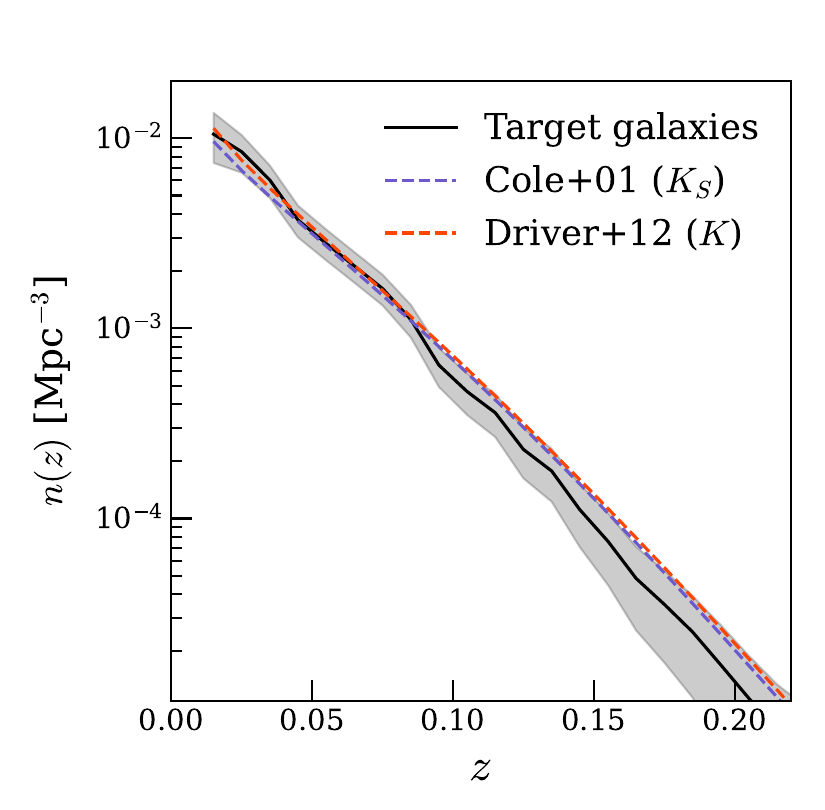}
\caption{Redshift distribution of galaxies with $K_S \leqslant 13.75$ calculated with the redshifts compiled from the literature. The gray shading indicates the $1\sigma$ range of Poisson error. The $n(z)$ from \citet{Cole2001} and \citet{Driver2012} are derived by integrating the best-fit Schechter functions. \label{fig:Fig3_z_nz}}
\end{figure}

To estimate the redshift distribution of the target galaxies, we use the redshifts collected from the literature and correct for the spectroscopic incompleteness. To minimize this correction, we retain only those HEALPix pixels whose completeness exceeds $0.5$. Galaxies are counted in each pixel and redshift bins of width $\Delta z = 0.01$. The redshift distribution, $n(z)$, is then calculated as follows. For the $i$th redshift bin and $j$th HEALPix pixel,
\begin{equation}\label{eq:1}
    n(z) = \frac{\sum_{j} \left(\frac{N_{ij}}{C_j}\right)}{\frac{4 \pi}{3} (r_{i+1}^{3} - r_i^{3})f_{C>0.5}}\,,
\end{equation}
where $N_{ij}$ is the number of galaxies in two-dimensional bins, $r_i$ is the comoving distance corresponding to the $i$th redshift bin, and $f_{C>0.5} \approx 0.27$ is the sky fraction covered by pixels with completeness greater than $0.5$. Figure \ref{fig:Fig3_z_nz} shows the estimated redshift distribution of our galaxy sample together with the distribution predicted from the measured $K_S$- and $K$-band luminosity functions in \citet{Cole2001} and \citet{Driver2012}. The median redshift is approximately $0.07$, and roughly 98\% of the galaxies lie at $z<0.2$. 

\subsection{Training Set: The IllustrisTNG Simulation \label{sec:Sec2.2}}
We use the IllustrisTNG (hereafter TNG) simulation \citep{Marinacci2018, Naiman2018, Nelson2018, Nelson2019, Pillepich2018, Pillepich2019, Springel2018} as our training set to learn the relation between DM subhalo and galaxy baryonic properties. The TNG simulation is a suite of magnetohydrodynamical simulations conducted with the moving-mesh code AREPO \citep{Springel2010}. The simulations adopt the cosmology from the Planck2015 results \citep{Planck2016} with $\Omega_{m,0}=0.3089$, $\Omega_{b,0}=0.0486$, $\Omega_{\Lambda,0}=0.6911$, $h=0.6774$, $\sigma_8=0.8159$, and $n_s=0.9667$, where $H_0=100\,h\,\rm{km}\,\rm{s}^{-1}\,\rm{Mpc}^{-1}$. DM halos are identified with the friends-of-friends (FOF) algorithm, and gravitationally bound substructures within the FOF halos are identified with the SUBFIND subhalo finder \citep{Springel2001}. SUBFIND identifies self-bound overdense structures by constructing a density field using the member particles of FOF halos. The galaxies identified in the TNG simulation reproduce the observed stellar-mass function \citep{Pillepich2018} and the stellar-mass-dependent clustering \citep{Springel2018} when an appropriate aperture is applied to galaxies. Thus, our ML model will become a suitable tool for constructing a mock catalog if it can learn the galaxy–halo relations encoded in the simulation.

Among the suite of the TNG simulations, we mainly employ TNG100-1, the highest-resolution version of the $75\, h^{-1}\,\rm{Mpc}$ box. TNG100-1 strikes a balance between statistics and resolution compared with TNG50 and TNG300. The run contains $1820^3$ DM particles with masses of $7.5\times10^6\,M_{\odot}$ and $1820^3$ initial gas cells with target baryon masses of $1.4\times10^6\,M_{\odot}$. This allows DM subhalos with as few as $100$ particles to be resolved down to $7.5\times10^8\,M_{\odot}$, while baryonic masses are resolved down to $1.4\times10^8\,M_{\odot}$. A stellar mass of $1.4\times10^8\,M_{\odot}$ roughly corresponds to $M_{K_S}\approx-17$ if we assume a stellar mass-to-light ratio of $\approx1 \,M_\odot/ L_\odot$ \citep{Bell2001, Bell2003} ($M_{K_S}\approx-19.3$ in the case of the lower-resolution run TNG300-1).

However, owing to the small volume $(75\,h^{-1}\,\rm{Mpc})^3$, the abundance of the rarest objects---galaxy group/cluster-scale halos---is limited in the TNG100-1 simulation. For example, at $z=0$, only 149 (10) subhalos have masses exceeding $10^{13}\,h^{-1}\,M_{\odot}$ ($10^{14}\,h^{-1}\,M_{\odot}$). Therefore, we additionally include subhalos with $M_{\rm{subhalo}}>10^{12}\,h^{-1}\,M_{\odot}$ from the TNG300-1 simulation to explore the possibility of compensating for this deficit. Nonetheless, simply combining the TNG300-1 subhalos is not straightforward, because their baryonic properties deviate systematically from those in the TNG100-1 simulation due to the lower resolution; the IllustrisTNG model was calibrated using a suite of $25\,h^{-1}\,\rm{Mpc}$ boxes at a mass resolution equivalent to that of TNG100-1. It has been shown in \citet{Pillepich2018} that simple scaling with a constant (${\sim}1.4$) works well for stellar mass in the high-mass regime $M_{\rm{star}} > 10^{10.5}\,M_{\odot}$. Thus, we apply a constant scaling factor to TNG300-1 baryonic properties when necessary. While constant scaling may not be a complete solution, we consider this experiment as a testbed, and do not pursue further fine-tuning. In Appendix \ref{AppA}, we show the scaling result between TNG300-1 and TNG100-1. In summary, we found a scaling factor of $1.44$ for stellar mass and luminosities, similar to \citet{Pillepich2018}, and factors of $0.83$, $1.19$, and $1.39$ for gas mass, star formation rate (SFR), and gas metallicity, respectively.

We restrict training to the subhalos in the $z=0$ snapshot, as the median redshift of A-SPEC is $0.07$. Instead, we make slight adjustments to compensate for the observational and evolutionary effects \citep[K and evolutionary (k+e) correction;][]{Bell2003} when constructing a mock catalog. In principle, MSSM can be trained with any subhalo catalog at an arbitrary epoch, and one can also apply that model to account for the evolutionary effect. 

\section{Machine Learning Strategies and Simulation Setup \label{sec:Sec3}}
In this section, we describe our ML model and strategies, including the input and output features and the metrics used to evaluate the accuracy of the model. We also describe how we design the $N$-body simulation to which the ML model is eventually applied.

\subsection{Machine-assisted Semi-Simulation Model \label{sec:Sec3.1}}
MSSM is an extremely randomized tree (ERT) regressor trained to predict the baryonic properties of subhalos from their DM properties in hydrodynamical simulations. It takes the DM properties of subhalos, along with their environmental and historical characteristics, as the input to predict baryonic properties of galaxies. The machine is composed of two-stage learning, where magnitudes in eight photometric bands ($M_U$, $M_B$, $M_V$, $M_K$, $M_g$, $M_r$, $M_i$, $M_z$) are first predicted and then added to the input to ultimately predict the remaining baryonic properties. In addition to the two-stage learning, the use of logged values and the inclusion of historical and environmental parameters enhance the predictive performance of the model. It has been shown in JK19 that MSSM is capable of predicting stellar mass ($M_{\rm{star}}$), gas mass ($M_{\rm{gas}}$), SFR, and gas metallicity ($Z_{\rm{gas}}$) with a mean binned error $\rm{(MBE)}=0.0013, 0.0010, 1.00, 0.010$ (see JK19 for a definition of MBE) in the range of baryonic properties they investigated. Moreover, the model was applied to the gigaparsec scale $N$-body simulation MDPL2 \citep{Knebe2018} and showed a comparable performance to the SAM \citep{Cora2018} with much less time and fewer resources. 

To train MSSM and apply it to an $N$-body simulation, we follow the prescription provided in JK19\footnote{We use the \texttt{ExtraTreesRegressor} in the \texttt{scikit-learn} library \citep{scikit-learn}.}, but slightly modify or add several features to potentially improve the performance of the machine. Below, we list the input and output features used for the regressor. Unless stated otherwise, subhalo or galaxy properties follow the definition of the TNG simulation, or equivalently, those of the SUBFIND subhalo finder.

The input DM features are as follows:
\begin{enumerate}[itemsep=0pt]
    \item $M_{\rm{subhalo}}$. The total mass of the subhalo.
    \item $v_{\rm{circ}}$. The maximum velocity of the spherically averaged rotation curve.
    \item $\sigma_v$. The velocity dispersion.
    \item $\boldsymbol{J}$. spin vectors measured for ($x$, $y$, $z$) axes.
    \item \textit{Subhalo-based environment}. We use four different definitions of subhalo environments as defined in JK19. In that study, environmental parameters are measured with neighboring subhalos contained within a $V_{\rm{box}}=(2\,\rm{Mpc})^3$ cube centered on the target subhalo. The parameters are as follows:
    \begin{enumerate}[label=(\roman*)]
        \item Local density $\sum M_i$/$V_{\rm{box}}$
        \item Number of local subhalos whose mass is larger than 80\% of the target subhalo
        \item Sum of masses over distance $\sum M_i$/$R_i$
        \item Mass over distance for the most massive subhalo in the box
    \end{enumerate}
    where $M_i$ and $R_i$ are the mass and distance to the neighboring subhalo. Here, we further impose a mass threshold to neighboring subhalos to match the mass resolution of the $N$-body simulation to which the MSSM is applied. Because typical $N$-body simulations have a lower resolution than TNG, a $(2\,\rm{Mpc})^3$ cube can contain only a small number of neighbors after the mass cut, especially in underdense regions. For example, in TNG100-1, the median number of subhalos in a $(2\,\rm{Mpc})^3$ cube is four (10) when counting only subhalos more massive than $10^{11}~(10^{10.5})\,h^{-1}\,M_{\odot}$. For this reason, apart from the number of local subhalos, we define the environmental volume by the distance to the $k$th nearest neighbor ($k$NN) and measure the remaining environmental parameters within that radius. We also include the distance to the $k$NN to the environment indicator. We adopt conventional values, $k\in \{ 1,2,3,4,5,8,15,20 \}$, and assess their impact on the machine.

    \item \textit{Subhalo anisotropy parameters} : We decide not to include the halo merger information originally suggested in JK19 due to the limited computational resources to construct merger trees in our $N$-body simulation. Instead, we add five subhalo properties that might assist the machine performance. The properties are as follows:
    \begin{enumerate}[label=(\roman*)]
        \item Bullock spin parameter \citep{Bullock2001}: $\lambda' \equiv J / \sqrt{2}MVR$
        \item $X_{\rm{off}}$: position offset between center and center of mass. The center of a subhalo is defined by the position of the particle with the minimum gravitational potential energy
        \item $V_{\rm{off}}$: velocity offset between the center and center of mass. We define the central velocity by averaging over the innermost 10\% of particles
        \item Axis ratio: $a/c$, $b/c$ where $a$, $b$, and $c$ are the lengths of major, intermediate, and minor axes computed with eigenvalues of the mass inertia tensor $I_{\alpha \beta}= \sum_{i} m_i x_{i,\alpha} x_{i,\beta} / \sum_i m_i$ 
    \end{enumerate}    
    These parameters are likely to embed signatures of recent mergers \citep{Wang2025} and are thus expected to partially compensate for the loss of merger tree information, as well as providing extra information on the galaxy--halo connection. 

    \item \textit{Particle-based environment}. As our training and application datasets are both from simulations, we consider additional environmental parameters based on simulation particles. The environment of a galaxy or a subhalo can be defined either isotropically (e.g., measured within a sphere) or anisotropically (e.g., distance from a nearby filament). To account for both, we add two extra environmental parameters measured from a smoothed density field and test their impact on the predictive performance of the model. To construct a density contrast field $\delta(\boldsymbol{r})$, we assign particles to a $0.75\,h^{-1}\,\rm{Mpc}$ resolution grid using a cloud-in-cell (CIC) scheme \citep{Hockney1988} and divide it by the mean number density. The density contrast field is then smoothed with a Gaussian filter with scale $R$ ($W(kR)=\exp[-(kR)^2 / 2]$) in Fourier space. From the smoothed field, we compute two quantities, $\delta_R$ and $\alpha_R$, at the location of subhalos using CIC interpolation. The $\delta_R$ is simply overdensity, and $\alpha_R$ is the tidal anisotropy parameter suggested in \citet{Paranjape2018}. They showed that $\alpha_R$ measured at a certain radius ($R=4R_{200b}$\footnote{$R_{200b}$ is the radius where the enclosed density is 200 times the background density.}) serves as an effective proxy for the tidal anisotropy around halos and is correlated with halo bias. The definition is as follows:
    \begin{align}
        \alpha_{R} &\equiv (1+\delta_{R})^{-1}\sqrt{q_{R}^2}\,,\\
        q_{R}^2 &= \frac{1}{2} \left[ (\lambda_3 - \lambda_1)^2 + (\lambda_3 - \lambda_2)^2 + (\lambda_2 - \lambda_1)^2 \right],
    \end{align}
    where $q_{R}^2$ is the tidal shear defined using the eigenvalues ($\lambda_1 \leqslant \lambda_2 \leqslant \lambda_3$) of the tidal tensor $T_{ij}$. The tidal tensor captures the local anisotropic deformation of matter induced by gravitational fields. The eigenvalues of the tidal tensor define the principal directions of collapse and expansion, and they are also used to classify different environments such as nodes, filaments, sheets, and voids. Tidal tensors are also computed in Fourier space as $T_{ij}(\boldsymbol{k})= (k_i k_j / k^2) \delta(\boldsymbol{k})$, and the corresponding tidal tensor grid in real space is obtained by inverse Fourier transformation. Although a length scale proportional to halo size was proposed, we use four fixed values of $R \in \{ 0.75, 1.5, 3, 5 \} \, h^{-1} \, \rm{Mpc}$ to probe the environment on different scales.

    \item \textit{Pseudomass accretion rate (pMAR)}. As merger tree information is not incorporated in this study, our input features cannot fully characterize the mass flow around galaxies. However, baryonic properties such as the SFR are closely associated with the halo mass accretion rate \citep{Dave2012, Lilly2013}. Therefore, we define a pMAR by tracing the most bound halo particle (MBP) of a subhalo. MBP--galaxy correspondence has been used in SAMs and empirical models to trace orphan galaxies \citep{Henriques2015, Hong2016}. We track the DM MBP identified at $z=0$ back to $z=0.2$ and measure the enclosed mass within spheres of radius $1$, $2$, $5$, and $10$ times the subhalo half-mass radius at $z=0$. A timescale of $\Delta t_{z=0.2}\simeq 2.5\,\rm{Gyr}$ is adopted to approximately follow mass flows on the halo dynamical timescale \citep[${\sim} 2.9\,\rm{Gyr}$;][]{Diemer2017}. The pMAR is defined as follows:
    \begin{equation}
        \rm{pMAR}(r) = \frac{\Delta M_{z=0.2}(r)}{M_{z=0}}\,,
    \end{equation}
    where $\Delta M_{z=0.2}(r)$ is the mass at $z=0$ ($M_{z=0}$) subtracted by that of $z=0.2$ for radius $r$. The pMAR can also have a negative value due to tidal stripping. As we use one epoch to estimate mass flow, we do not divide it by the time interval but just normalize to the mass at the later epoch. We also add $\delta_R$ measured at those radii to see whether an aperture proportional to subhalo size is more effective than the fixed aperture. In the following sections, we present the results after incorporating the pMAR into the \textit{particle-based environment}.

    \item \textit{Host halo properties} : We adopt two features related to the host FOF halo of a subhalo based on SUBFIND. The first one is a Boolean flag indicating whether a subhalo is the central subhalo. The second is the distance from the center of the host halo normalized by $R_{200b}$.

\end{enumerate}

The output baryonic features are as follows:
\begin{enumerate}[itemsep=0pt]
    \item $M_{band}$. Stellar magnitudes (Vega) in ($U$, $B$, $V$, $K$, $u$, $g$, $r$, $z$) photometric bands. Stellar magnitudes are computed by treating each stellar particle as a simple stellar population (SSP) \citep{Nelson2018}, with a spectrum assigned as a function of age and metallicity. The galaxy luminosity in each band is then obtained by convolving the summed spectrum with the corresponding passband. 
    \item $M_{star}$. Stellar mass
    \item $M_{gas}$. Gas mass
    \item $SFR_{100\,Myr}$. The SFR averaged over $100\,\rm{Myr}$. The timescale should ideally be determined based on the choice of star formation indicator in observations. However, as we do not compare the SFR with observational data in this work, we adopt an arbitrary value. Throughout the paper, we denote $\rm{SFR}_{100\,\rm{Myr}}$ simply as ``SFR,'' as no other averaging timescales are considered. 
    \item $Z_{\rm{gas}}$. Mass-weighted gas metallicity
\end{enumerate}
We apply a $30\,\rm{kpc}$ aperture to the baryonic properties to match the observed stellar-mass and luminosity functions. No aperture is imposed on gas mass, as we do not compare it with observations in this study. In practice, any aperture could be chosen to reflect specific observational strategies, such as fiber or beam sizes. 

A training sample is restricted to subhalos satisfying the conditions listed below to ensure reliable subhalo and galaxy properties. Our selection criteria are as follows:
\begin{enumerate}[label=(\arabic*), itemsep=0pt, topsep=8pt, parsep=0pt]
    \item $N_{\rm{DM}} \geqslant 40$ and $N_{\rm{star}} \geqslant 40$
    \item $\textrm{SFR} > 10^{-2.15}\,M_{\odot}\,\rm{yr}^{-1}$
    \item $M_{\rm{gas}} > 0$
    \item $Z_{\rm{gas}} > 0$
\end{enumerate}

where $N_{\rm{DM}}$ and $N_{\rm{star}}$ are the numbers of DM and stellar particles in a subhalo. Although a higher threshold (e.g., 100 particles) would better capture internal halo structure, we kept the cut as low as possible to minimize selection bias. Arbitrarily increasing one of the cuts would eliminate particular subhalo populations. It is worth noting that our training sample is incomplete in terms of stellar content for a given subhalo mass (i.e., no subhalos with $N_{\rm{DM}}\geqslant40$ and $N_{\rm{star}}<40$). The absence of these subhalos leads to an overestimation of the stellar mass for less massive subhalos in the application phase. However, the stellar-mass cut ($N_{\rm{star}}\geqslant40$) is approximately three magnitudes fainter than our requirement ($M_{K_S}\lesssim-19.5$) so that the galaxies of our interest are not affected. We have checked that lowering the stellar-mass cut does not alter the final mock product. While some subhalos exhibit low DM content relative to their stellar component \citep[e.g.,][]{MonteroDorta2024}, we ignore this population due to its scarcity. Conversely, we include all subhalos that contain at least one gas cell. Although this choice may reduce the reliability of gas-related quantities, quiescent galaxies can be intrinsically gas poor, and our $K_S$-selected sample is defined by stellar light. We also impose a resolution cut to the SFR corresponding to the 100 Myr timescale \citep{Donnari2019}. We exclude subhalos whose gas-related properties are set to ``0,'' because in reality (and in higher-resolution simulations) these quantities should be nonzero but very small. These unphysical placeholder values (0) could bias the model predictions. Except for stellar magnitudes, we take log values of all the subhalo properties. The evaluation metrics described below are also calculated using logged values.

Training and test sets are randomly divided into 80\% and 20\%. For the test set, we report three conventional metrics---the coefficient of determination $R^2$, mean absolute error (MAE), and Pearson correlation coefficient $\rho$---to quantify the accuracy of the model under different settings. The metrics are defined as follows:
\begin{align}
    R^2 &= 1 - \dfrac{\sum_{i=1}^{n} (y_i-y_{i,\rm{pred}})^2}{\sum_{i=1}^{n} (y_i-\overline{y})^2}\,, \\
    \rm{MAE} &= \dfrac{1}{n}\sum_{i=1}^{n} \left| y_i - y_{i,\rm{pred}} \right|\,, \\
    \rho &= \dfrac{\sum_{i} (y_i-\overline{y})(y_{i,\rm{pred}}-\overline{y}_{\rm{pred}})}{\sqrt{\sum_{i} (y_i-\overline{y})^2 \sum_{i}(y_{i,\rm{pred}}-\overline{y}_{\rm{pred}})^2}}\,,
\end{align}

\noindent where $y$ and $y_{\rm{pred}}$ are the true and predicted baryonic properties, respectively. We additionally assess the performance of the model by comparing each metric between the training and test sets to identify potential overfitting and evaluate the generalization capability. Furthermore, we perform fivefold cross validation (CV) with a fixed random seed, in which the training and test sets vary across folds. This procedure enables us to estimate the average generalization performance and quantify the variability of these metrics across different folds. We search for the best-performing hyperparameters using Optuna\footnote{\href{https://optuna.readthedocs.io/en/stable/}{https://optuna.readthedocs.io/en/stable/} \citep{optuna_2019}} by maximizing the overall $R^2$. However, we observe only marginal performance gains across hyperparameter settings; preventing overfitting by tuning \texttt{max\_depth}, \texttt{min\_samples\_leaf}, and \texttt{min\_samples\_split} remains important.

\subsection{$N$-body Simulation: NASIM \label{sec:Sec3.2}}

\begin{figure*}[t]
\centering
\includegraphics[width=0.99\textwidth]{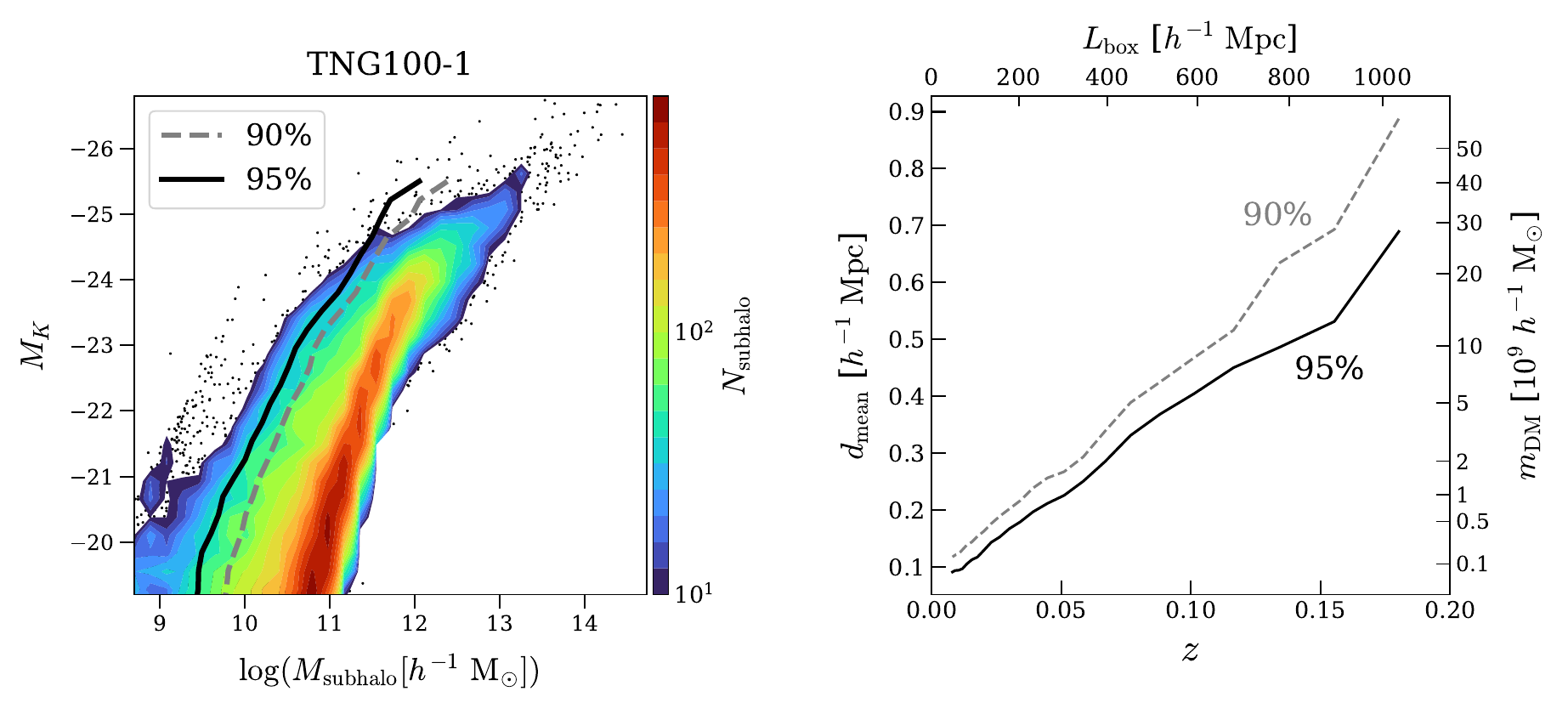}
\caption{Left: absolute magnitude $M_{K}$ as a function of subhalo mass for TNG100-1 subhalos. The gray and black dashed lines show the top 90\% and 95\% massive subhalos for a given $M_K$.  Right: the resolution required at each redshift ($z$) or box size ($L_{\rm{box}}$). $d_{\rm{mean}}$ denotes the mean particle separation, and $m_{\rm{DM}}$ denotes the mass of a DM particle. The redshift is converted to a box size by locating an observer at the center of the box. Minimum subhalo masses are converted to particle masses and mean particle separations, assuming 40 particles per subhalo. \label{fig:Fig4_TNG100_M_MK}}
\end{figure*}

As MSSM employs high-resolution hydrodynamical simulations, it assumes a subhalo--galaxy correspondence model. Therefore, the $N$-body simulation used to generate the DM field needs to resolve subhalos down to the mass that covers baryonic properties of interest. A-SPEC is a flux-limited survey in the $K_S$ band, namely the limiting absolute magnitude varies over $3\,\textrm{mag}$ depending on the redshift, as in Figure \ref{fig:Fig2_z_MKs}. To estimate the minimum subhalo mass corresponding to a given luminosity at each redshift, we utilize our training set, the TNG100-1 simulation. In the left panel of Figure \ref{fig:Fig4_TNG100_M_MK}, we show the distribution of subhalos from TNG100-1 in $(M_{\rm{subhalo}},M_{K})$ space. Based on this relation, we calculate the minimum subhalo mass required to cover $\gtrsim 90\,(95)\%$ of the subhalos at each $M_{K}$, as shown in gray (black) lines. Assuming that at least 40 particles are necessary to robustly identify a subhalo, the resolution limit as a function of redshift is shown in the right panel of Figure \ref{fig:Fig4_TNG100_M_MK}. The redshift can also be converted to a box size by placing an observer at the center of the simulation box. The requirements of the simulation can be summarized such that the box size ($x$-axis) and resolution ($y$-axis) are constrained to lie below the curves. For instance, to resolve the top 90\% heaviest subhalos at $z=0.01$ with limiting $M_{K_S}=-19.5$, a minimum subhalo mass of ${\sim} 6 \times 10^{9}$ $h^{-1}\,M_{\odot}$ is required (left panel), and thus a DM particle mass of $1.5\times10^8$ $h^{-1}\,M_{\odot}$ is needed for the $N$-body simulation (right panel). This corresponds to more than a trillion DM particles to cover our survey volume, which extends up to $z=0.2$.

Although such an approach is technically feasible given current advancements \citep{Potter2017, Heitmann2019, Ishiyama2021}, it remains computationally too expensive in terms of both resources and time, and offers limited benefit relative to the cost in this study. Instead, we carry out an $N$-body simulation tailored to the requirements of our flux-limited survey. Specifically, we perform a zoomed-in simulation NASIM\footnote{NASIM is an alternative version of A-SIM described in \citet{Dupuy2026}} composed of multiple zoomed-in regions with a progressively higher resolution corresponding to lower redshift. The zoomed-in regions should satisfy two conditions. First, the outermost box should extend beyond the redshift range ($z>$0.2) of the target galaxies. This corresponds to $L_{\rm{box,zoom}} \approx 1200$ $h^{-1}\,\rm{Mpc}$ when placing an observer at the center of the box. Second, the resolution of each box should be sufficient to capture all the target galaxies located between its boundaries and those of the next finer zoomed-in level.

\begin{deluxetable*}{ccccccc}
\tablecaption{$N$-body Simulation Setup \label{tab:Tab1}}

\tablehead{
\colhead{Simulation} &
\colhead{$L_{\rm{box}}$} &
\colhead{$N_p$} &
\colhead{$m_{\rm{DM}}$} &
\colhead{$d_{\rm{mean}}$} &
\colhead{$\epsilon$} &
\colhead{Redshift Range} \\
\colhead{} &
\colhead{($h^{-1}\,\rm{Mpc}$)} &
\colhead{} &
\colhead{($10^9$\,$h^{-1}\,M_{\odot}$)} &
\colhead{($h^{-1}\,\rm{Mpc}$)} &
\colhead{($h^{-1}\,\rm{kpc}$)} &
\colhead{}
}

\startdata
Periodic & 3072 & $1024^3$ & 2300 & 3       & 60     & \nodata \\
\tableline
\multicolumn{1}{c}{} &
\multicolumn{1}{c}{$L_{\rm{zoom}}$} &
\multicolumn{4}{c}{High-resolution particles} &
\multicolumn{1}{c}{} \\
\cline{2-2}\cline{3-6}
zoom1    & 1536 & $2048^3$ & 36   & 0.75     & 15     & 0.17--0.26 \\
zoom2    & 1050 & $2800^3$ & 4.5  & 0.375    & 7.5    & 0.08--0.17 \\
zoom3    & 525  & $2800^3$ & 0.56 & 0.1875   & 3.75   & 0.03--0.08 \\
zoom4    & 225  & $2400^3$ & 0.07 & 0.09375  & 1.875  & 0--0.03 \\
\enddata

\tablecomments{$L$ denotes the box size of each simulation box, $N_p$ and $m_{\rm{DM}}$ are the number and masses of DM particles, $d_{\rm{mean}}$ is the mean particle separation, and $\epsilon$ is the softening length. The redshift range covered by each simulation box is shown in the last column (see Figure \ref{fig:Fig5_zoomin_Config}). For zoomed-in simulations, we tabulate the values of the highest-resolution particles.}
\end{deluxetable*}

\begin{figure*}[t]
\centering
\includegraphics[width=0.99\textwidth]{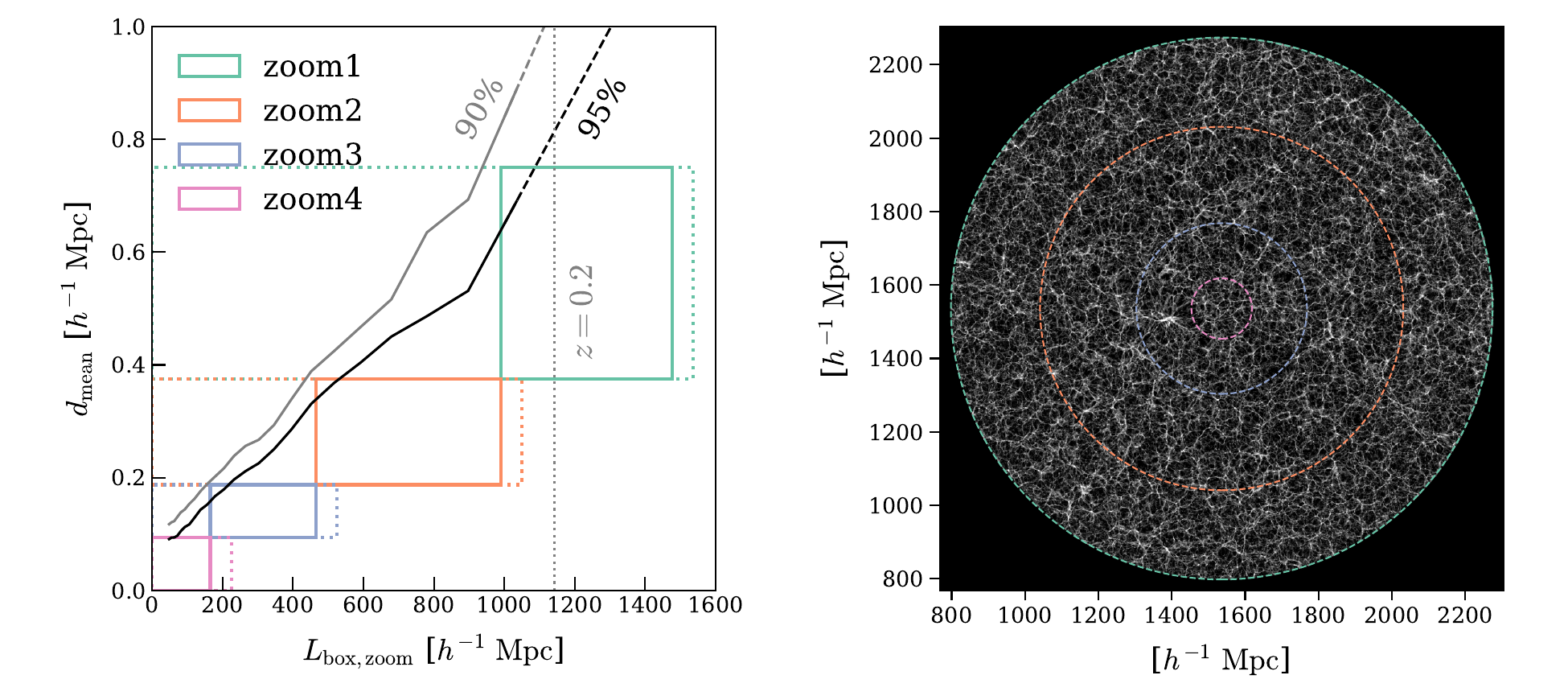}
\caption{Left: the sizes and resolution of zoomed-in boxes of a periodic box size $L_{\rm{box}}=3072\,h^{-1}\,\rm{Mpc}$. The gray and black lines are the same as the curves in Figure \ref{fig:Fig4_TNG100_M_MK} indicating the resolution requirements. The horizontal dotted lines denote the potential resolution expressed as the mean particle separation $L_{\rm{box}}/2^p$ where $p\in\{12,13,14,15\}$. Four different color boxes show each redshift range covered by zoomed-in boxes that lie under the curve. While the vertical dotted lines show the exact box size, the vertical solid lines denote the boundaries used for the mock data. Right: the density field ($3\,h^{-1}\,\rm{Mpc}$ slice) used for constructing the mock galaxy catalog. Particles located between the colored circles are extracted from each zoomed-in simulation.  \label{fig:Fig5_zoomin_Config}}
\end{figure*}

To generate zoomed-in initial conditions, we utilize the MUlti-Scale-Initial-Conditions (MUSIC) code \citep{Hahn2011}. In the MUSIC code, resolutions of the zoomed-in regions are determined by the periodic box size divided by powers of two. Thus, for a given periodic box size, zoomed-in box sizes and resolutions are fully specified to keep the box under the curves in the right panel of Figure \ref{fig:Fig4_TNG100_M_MK}. We explore periodic box sizes in the range of $[2000, 4000]$ $h^{-1}\,\rm{Mpc}$ and find that $L_{\rm{box}}\sim 3000$ $h^{-1}\,\rm{Mpc}$ with four zoomed-in regions has the minimum total number of particles. Possible resolutions are determined by $L_{\rm{box}}$/$2^p$, where $p\in\{12,13,14,15\}$ and the box sizes are constrained by the resolution requirements. The final setup of the simulations is shown in Table \ref{tab:Tab1} and in the left panel of Figure \ref{fig:Fig5_zoomin_Config}. We add a $\gtrsim 30\,h^{-1}\,\rm{Mpc}$ buffer to each zoomed-in box (i.e., box sizes are increased by 60 $h^{-1}\,\rm{Mpc}$) to minimize boundary effects and compensate for the displacement of particles at $z=0$. The right panel of Figure \ref{fig:Fig5_zoomin_Config} shows the $3\,h^{-1}\,\rm{Mpc}$ slice of the density field from NASIM, which is used to construct the mock galaxy catalog.

The zoomed-in simulations are conducted using the publicly available version of the GADGET-2 code \citep{Springel2005} with Planck2015 cosmology for consistency with the TNG simulations. The particles are evolved from the initial conditions generated at $z=127$ using the second-order Lagrangian perturbation theory (2LPT) implemented in the MUSIC code. We note that the IllustrisTNG initial conditions are generated with the Zel'dovich approximation (ZA), while ours use 2LPT. However, both simulations are initialized at $z=127$. At such a high starting redshift, the residual transients from the ZA--2LPT difference should be at most a few percent by $z=0$ \citep[e.g.,][]{Crocce2006,Knebe2009,LHuillier2014}, well within the intrinsic scatter of the ML halo--galaxy mapping.
The softening length is set to be 1/50 of the mean particle separation. In the application phase, MSSM takes subhalo properties such as $v_{\rm{circ}}$, $\sigma_v$, and spin as inputs. These measurements depend on the resolved inner mass distribution, which can vary with the gravitational softening length. Therefore, the difference in softening lengths between the $N$-body and TNG simulations may lead to systematic differences in these inputs and propagate into the MSSM inference. The impact of softening length on halo properties has been investigated in \citet{Zhang2019} and \citet{Mansfield2021}, but fine-tuning across different resolutions is not straightforward. We adopt a conventional value for a large-scale cosmological simulation and make adjustments by comparing the distribution functions of subhalo properties across simulations.\\

\section{Results \label{sec:Sec4}}

\subsection{MSSM Training Result \label{sec:Sec4.1}}
In this section, we demonstrate the training result of MSSM. We examine the impact of additional input features that we adopted in Section \ref{sec:Sec3.1} and assess the overall performance of our machine.

\begin{figure*}[htbp]
\centering
\includegraphics[width=0.7\textwidth]{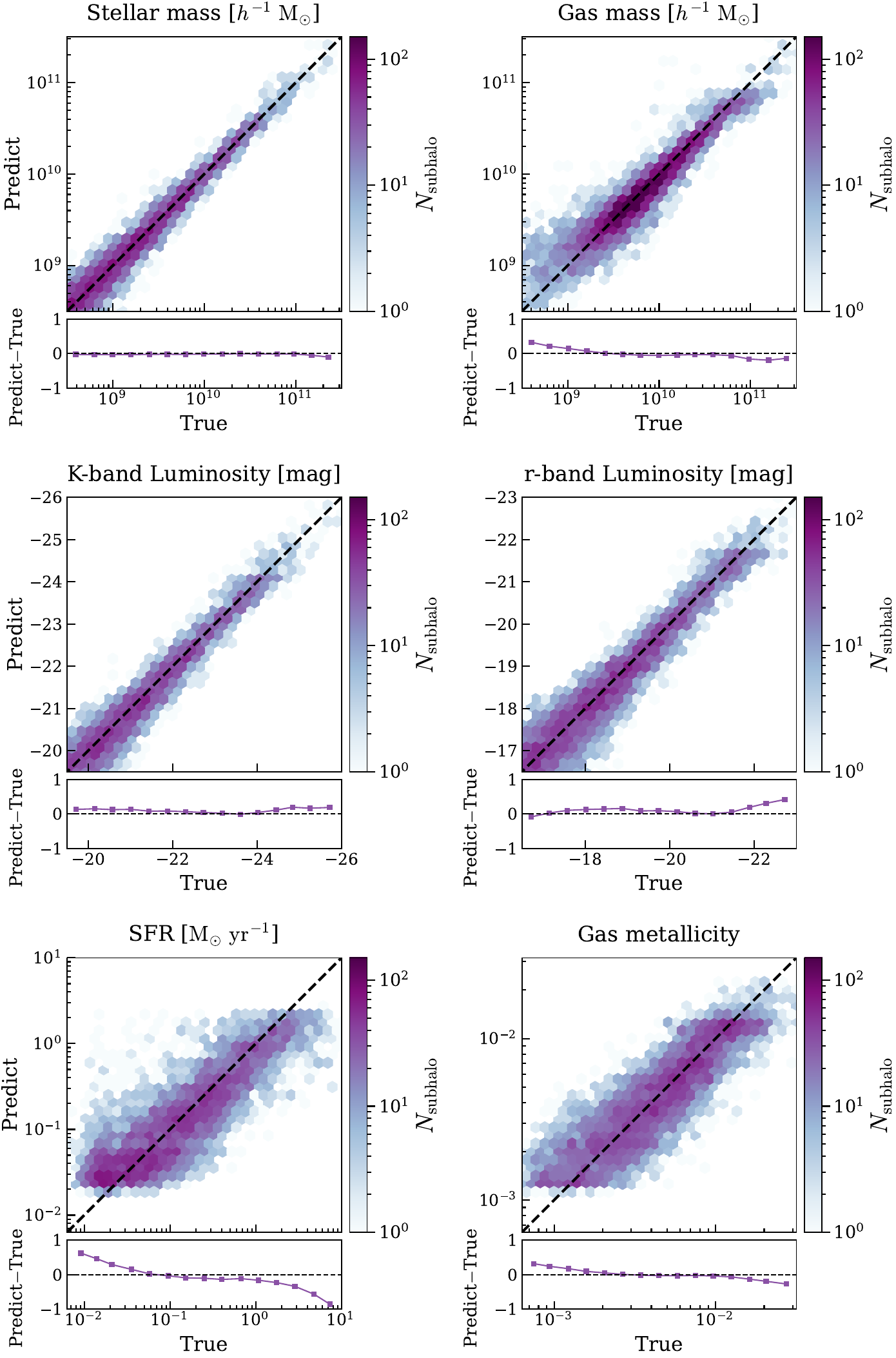}
\caption{Top: one-to-one comparison between true and predicted baryonic properties of the test dataset. The model is trained with TNG100-1 (without TNG300-1: see the blue lines of Figure \ref{fig:Fig7_Fiducial_Predict_Phi}). The color map shows the number of subhalos in each bin. Bottom: the difference between mean binned values. Star-related quantities show reasonable agreement, but $M_{\rm{gas}}$, SFR, and $Z_{\rm{gas}}$ show a difference in the tails of the distributions.
\label{fig:Fig6_Fiducial_Predict_OnetoOne}}
\end{figure*}

\begin{figure*}[htbp]
\centering
\includegraphics[width=0.7\textwidth]{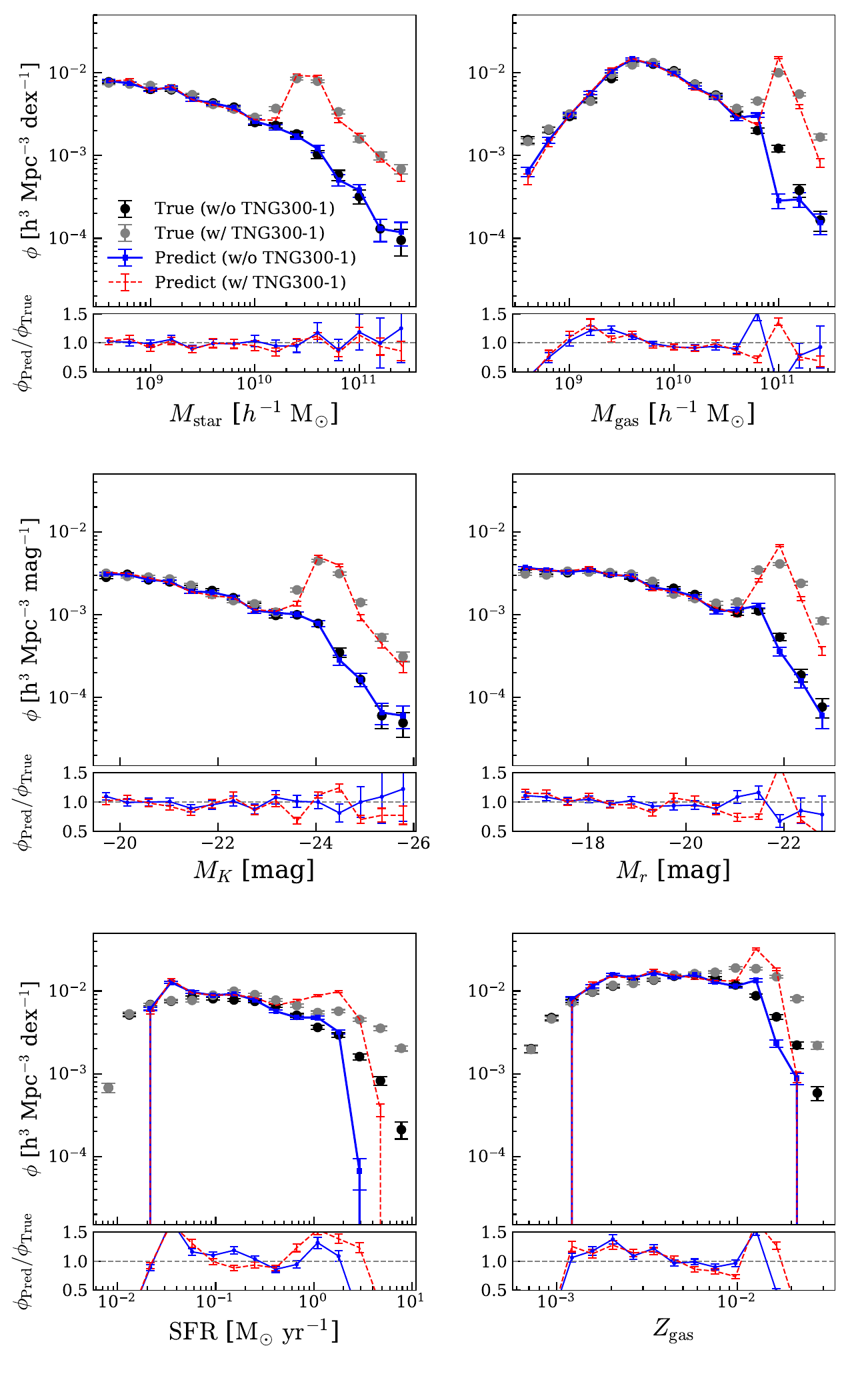}
\caption{Top: distribution functions, $\phi$, of six predicted baryonic properties for the test dataset. Bottom: blue lines show the ratio of predicted $\phi$ to the true values from TNG100-1, while red lines show the same ratios when subhalos more massive than $10^{12}\,h^{-1}\,M_{\odot}$ from TNG300-1 are included in the training set. There is no significant improvement in predictive performance at the tails of the distributions, even when the massive subhalos are taken into account. 
\label{fig:Fig7_Fiducial_Predict_Phi}}
\end{figure*}

Comparisons between true and predicted baryonic properties in the test dataset are shown in the upper panels of Figures \ref{fig:Fig6_Fiducial_Predict_OnetoOne} and \ref{fig:Fig7_Fiducial_Predict_Phi} for the fiducial model, illustrating the one-to-one relations in the former and distribution functions $\phi$ in the latter. The fiducial model uses only the input features adopted in the original work by JK19 ($M_{\rm{subhalo}}, v_{\rm{circ}},\sigma_v, \boldsymbol{J}$, and \textit{subhalo-based environment}). The bottom panels show the ratio between true and predicted values for mean binned statistics. In the demonstrated model, the mass cut applied to neighboring subhalos is set to match the resolution of the zoom3 simulation ($M_{\rm{subhalo}}>2.3\times10^{10}\,h^{-1}\,M_{\odot}$), which corresponds to the median redshift ($z=0.07$) of the target galaxies. However, we note that the contribution from the environmental parameters has a secondary effect, and the overall trends do not strongly depend on the mass cut.
Both figures indicate a similar trend, where the star-related properties (stellar mass and luminosities) show close agreement within $0.2\,\rm{dex}$, whereas gas-related properties (gas mass, SFR, and $Z_{\rm{gas}}$) exhibit similar deviations from the true values, being systematically underestimated at the high-end tails and overestimated at the low-end tails. In particular, the predictive performance of the SFR and $Z_{\rm{gas}}$ is substantially inaccurate compared to other baryonic properties. The model fails to predict the low and high ends of the distribution; galaxies with high and low SFRs do not appear in the prediction. The three metrics that we use to assess our model are summarized in Table \ref{tab:Tab2}. As shown in the figures, our model reaches $R^2$ over 0.9 for stellar mass and most of the stellar luminosities. $M_{\rm{gas}}$ also reaches $R^2=0.89$ and $\rm{MAE}=0.13\,\rm{dex}$ even though it deviates from the true values systematically, as noted above. One noticeable trend is that the stellar luminosity prediction worsens as the wavelength becomes shorter (e.g., compare the $K$ band and $U$ band), already indicating that star formation activity will not be well captured by the input features. 

\begin{deluxetable}{lccc}
\tablecaption{Predictive Performance of the Fiducial Model \label{tab:Tab2}}
\tablewidth{0pt}
\tabletypesize{\small}

\tablehead{
\colhead{} & \multicolumn{3}{c}{Zoom3} \\
\cline{2-4}
\colhead{Quantity} & \colhead{$R^2$} & \colhead{MAE} & \colhead{$\rho$}
}

\startdata
$M_{\rm{star}}$ & 0.955 (0.001) & 0.123 (0.001) & 0.977 (0.001) \\
$M_{\rm{gas}}$  & 0.887 (0.004) & 0.132 (0.001) & 0.943 (0.002) \\
SFR             & 0.697 (0.004) & 0.274 (0.002) & 0.835 (0.002) \\
$Z_{\rm{gas}}$  & 0.771 (0.002) & 0.136 (0.000) & 0.878 (0.001) \\
$M_U$           & 0.846 (0.003) & 0.516 (0.003) & 0.920 (0.002) \\
$M_B$           & 0.890 (0.003) & 0.433 (0.003) & 0.943 (0.001) \\
$M_V$           & 0.915 (0.002) & 0.384 (0.003) & 0.957 (0.001) \\
$M_K$           & 0.937 (0.001) & 0.361 (0.003) & 0.968 (0.001) \\
$M_g$           & 0.900 (0.002) & 0.414 (0.003) & 0.949 (0.001) \\
$M_r$           & 0.922 (0.002) & 0.370 (0.003) & 0.960 (0.001) \\
$M_i$           & 0.929 (0.002) & 0.358 (0.003) & 0.964 (0.001) \\
$M_z$           & 0.933 (0.002) & 0.353 (0.003) & 0.966 (0.001) \\
\enddata

\tablecomments{Subhalos resolved in the zoom3 simulation are considered when calculating the subhalo environment using the nearest neighbors. The metrics show the mean of fivefold cross validation and the numbers in the parentheses are the standard deviations.}
\end{deluxetable}

\begin{figure}[t]
\centering
\includegraphics[width=0.99\linewidth]{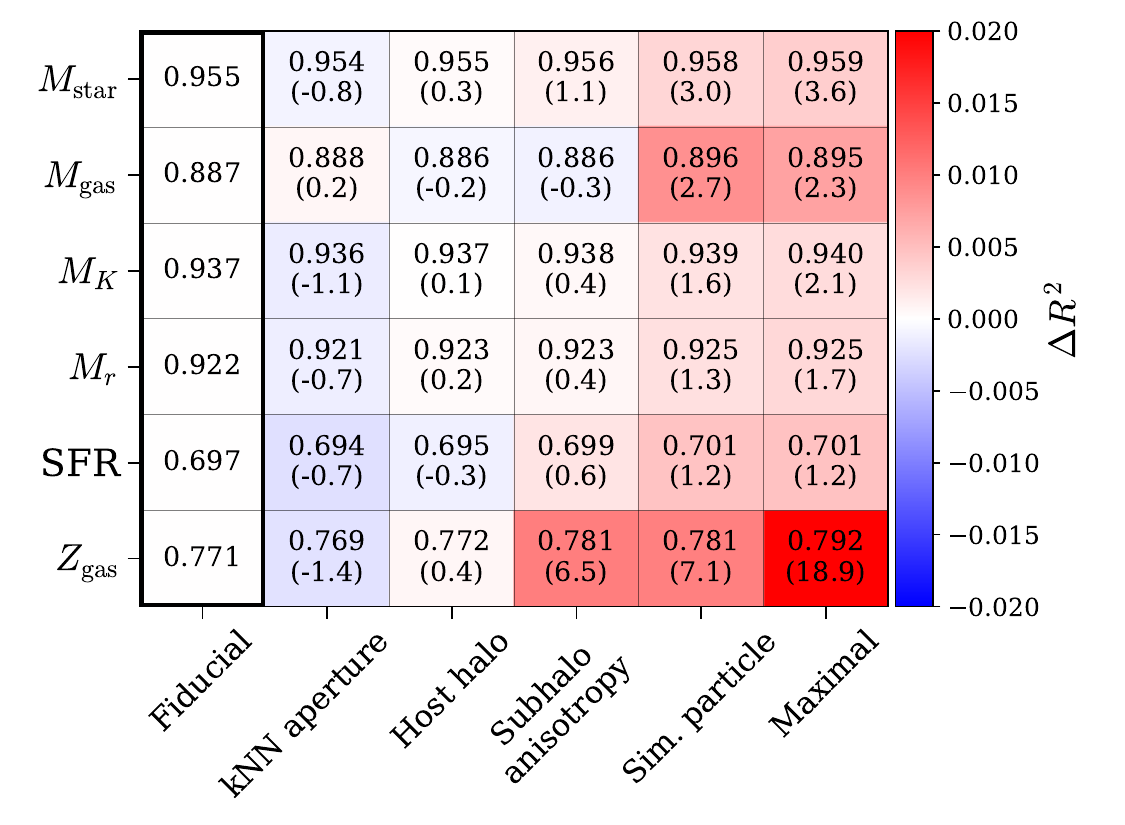}
\caption{$R^2$ of six key predicted baryonic properties for different models. The ``kNN aperture'' refers to the case when the aperture for environmental parameter measurement is defined by the nearest neighbors (see \textit{subhalo-based environment} in Section \ref{sec:Sec3.1}). The ``Sim. particle'' denotes the \textit{particle-based environment}. The ``Maximal'' denotes the case when all the input features in the left columns are included. Improvements relative to the fiducial model ($\Delta R^2$) are shown as a color map in the background, and the numbers in the parentheses are the significance of the enhancement, calculated as $\Delta R^2$ divided by the fivefold CV standard deviation. Adopting subhalo anisotropy parameters and the environment measured with simulation particles improves the model, especially for $Z_{\rm{gas}}$.
\label{fig:Fig8_MSSM_zoom3_summary}}
\end{figure}

Our sample selection criteria, which exclude gas-poor and low-SFR subhalos, are likely to affect the result as they skew the distribution function of our sample. We further discuss the impact of excluded subhalos in Section \ref{sec:Sec5.1.2}. Apart from this selection effect, another caveat of our training sample is the scarcity of massive subhalos in the TNG100-1 simulation, as mentioned in Section \ref{sec:Sec2.2}. To investigate the potential impact of this limitation, we use the augmented training sample constructed with TNG300-1, scaled to reproduce the distribution functions  of TNG100-1. The prediction results obtained from this combined sample are shown as red dashed lines in Figure \ref{fig:Fig7_Fiducial_Predict_Phi}. Some of the quantities such as stellar and gas masses are slightly improved, but the improvement is not significant considering the error bars. Even with the inclusion of a larger number of massive subhalos, the machine still fails to reproduce the population of massive subhalos, indicating that the limited sample size is not the primary cause of the performance degradation. However, we note that the distribution function of the augmented sample is nonphysical, and thus the impact of sample size should be further investigated.

Figure \ref{fig:Fig8_MSSM_zoom3_summary} shows the $R^2$ values of the six key baryonic properties as additional feature groups are added to the fiducial model. The difference in $R^2$ relative to the fiducial model ($\Delta R^2$) is color coded. To quantify the significance of the improvement, we divide the $\Delta R^2$ by the fivefold CV standard deviation of each model, which are the numbers in the parentheses. The absolute increase in $R^2$ is small, less than $0.02$ in the best case. Including the \textit{particle-based environment} shows an overall improvement compared to the fiducial model, especially in the case of $Z_{\rm{gas}}$. Including subhalo anisotropy parameters also improves the model overall, but only marginally considering the CV standard deviation. The inclusion of the $k$NN aperture and host halo properties did not improve the model significantly. However, in the case of the maximal model, all of the predictions increased. In this figure, we summarize the relative accuracy when groups of features are added to the model; the importance of the features is further discussed in Sections \ref{sec:Sec5.1.1} and \ref{sec:Sec5.2}. A table containing the full metric information of $R_{\rm{train}}^2$, MAE, $\rho$, and CV standard deviations is provided in Appendix \ref{AppB}.

\subsection{Basic Validation of the $N$-body Simulation \label{sec:Sec4.2}}
In this section, we summarize the basic results from the four zoomed-in $N$-body simulations.
We compare the results with the higher-resolution runs of our zoomed-in simulation (e.g., zoom1 with zoom2) and the TNG100-1 simulation, focusing on the mass function and mass-dependent clustering of subhalos. We consider both DM and baryon particles/cells in the TNG100-1 simulation when calculating the mass. DM halos and subhalos at $z=0$ in NASIM are identified using the publicly available version of SUBFIND \citep{Springel2021} to minimize the dependence on the halo finder.\footnote{Various halo finders adopt distinct algorithms for identifying substructures, as well as for membership selection and unbinding. Differences in halo properties arising from the choice of halo finder can be mitigated through postprocessing, but in this work we simply use the same halo finder (see \citealp{Knebe2011} and \citealp{Onions2012} for a detailed discussion).}

\subsubsection{Halo and Subhalo Mass Function \label{sec:Sec4.2.1}}
\begin{figure*}[t]
\centering
\includegraphics[width=0.8\textwidth]{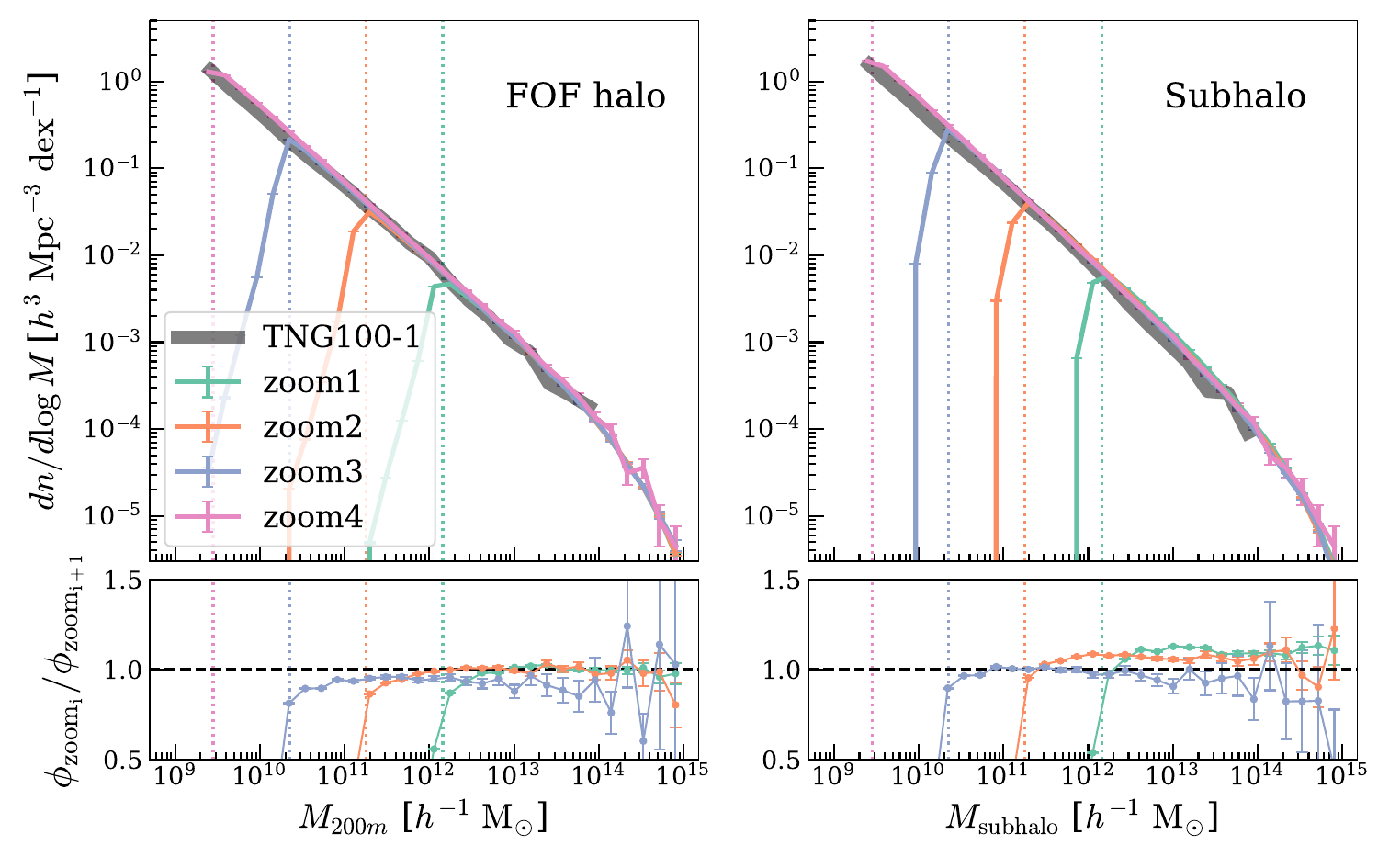}
\caption{Left: mass functions of FOF halos from the four zoomed-in simulations. Halo mass is defined as $M_{200\rm{m}}$, the mass enclosed within a spherical overdensity of 200 times the mean matter density. Right: the same as the left panel, but for subhalos identified using SUBFIND. Both results are comparable to those from higher-resolution runs within 10\%. In both panels, vertical dotted lines indicate the mass corresponding to 40 particles in each simulation, and the thick gray lines denote the measurements from TNG100-1. Our results are consistent with those from TNG100-1 within 30\%.
\label{fig:Fig9_NASIM_MassFunc}}
\end{figure*}

We use the $M_{200\rm{m}}$ of FOF halos, namely the mass enclosed within the radius at which the density reaches 200 times the mean matter density, $\rho_{m}=\frac{3 H_{0}^2}{8 \pi G} \Omega_{m,0}$. The subhalo mass is defined by the sum of the entire member particles bound to the subhalo. We exclude the $30\,h^{-1}\,\rm{Mpc}$ region (see Section \ref{sec:Sec3.2}) near the boundary when measuring statistics to avoid contamination due to frequent interactions with low-resolution particles\footnote{We find that, at $z=0$, the low-resolution particles are located at least $10\,h^{-1}\,\rm{Mpc}$ from the boundary.}. In Figure \ref{fig:Fig9_NASIM_MassFunc}, we show the abundance of halos and subhalos in our four zoomed-in simulations. Both of the mass functions are comparable to the higher-resolution runs within 10\% for objects resolved with $\geqslant40$ DM particles. A residual systematic offset in the subhalo mass function arises because a higher resolution identifies smaller substructures. The corresponding measurements from the TNG100-1 simulation are overplotted as gray lines, and both mass functions agree with them to within $30\%$ ($0.1\,\rm{dex}$).

\begin{figure*}[t]
\centering
\includegraphics[width=\textwidth]{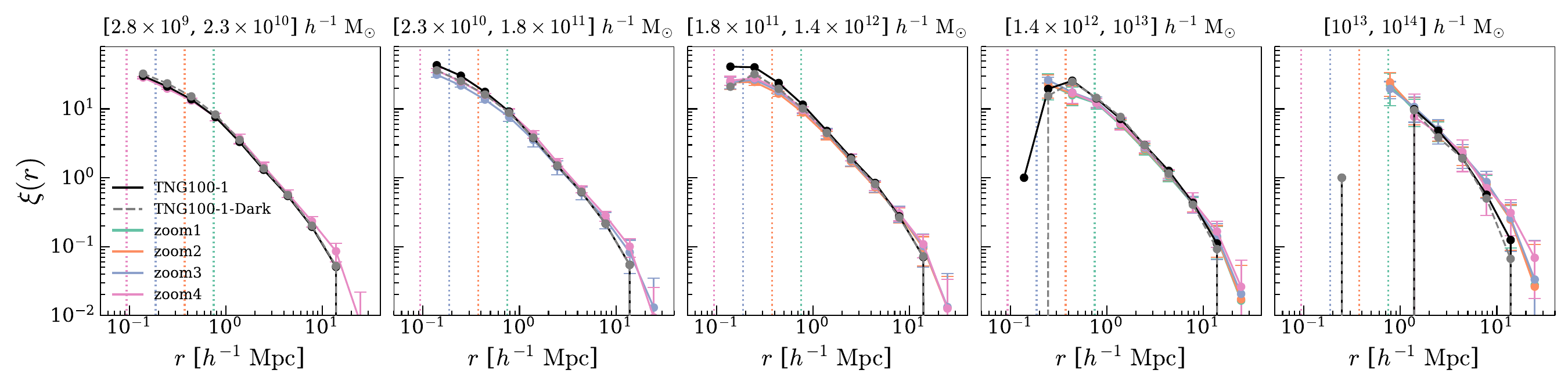}
\caption{Mass-dependent two-point correlation functions of subhalos in TNG100-1 and zoomed-in simulations measured in real space. We also show the results from the DM counterpart simulation (TNG100-1-Dark) for comparison (gray dashed line). Mass bins are set by the 40 DM particles of a zoomed-in simulation and the next finer zoomed-in level. Vertical dotted lines denote the mean particle separation of each simulation. The uncertainties in the zoomed-in simulations are estimated by partitioning the box by the TNG100-1 box size $(75\,h^{-1}\,\rm{Mpc})^3$. They are consistent within uncertainties.
\label{fig:Fig10_NASIM_subhalo_xi}}
\end{figure*}

Following this comparison of mass functions, we also examine subhalo size and its correlations. As mentioned in Section \ref{sec:Sec3.2}, MSSM application results depend on the sizes of subhalos. In Appendix \ref{AppC}, we show the distributions of properties that are related to the sizes of subhalos across different simulations---subhalo half-mass radius, $v_{\rm{circ}}$, $\sigma_v$, and spin in our zoomed-in simulations and TNG100-1. In summary, they are consistent with TNG100-1-Dark to within $0.3\,\rm{dex}$, but relative to TNG100-1 they exhibit a noticeable shape difference: a kink.

\subsubsection{Clustering of Subhalos \label{sec:Sec4.2.2}}
The ultimate goal of our mock galaxy catalog is to reproduce both the one-point and two-point statistics of the target galaxies. However, MSSM is trained on the relationship between DM and baryonic properties (i.e., one-point functions) of subhalos and does not encode their spatial distribution. As a result, the two-point statistics of the mock data may deviate from those of TNG100-1 even if the MSSM perfectly assigns baryonic properties to DM subhalos. Thus, before we delve into the application phase, clustering of DM subhalos should be guaranteed. To validate this, we compare the mass-dependent clustering of subhalos in our zoomed-in simulations with that in TNG100-1. Subhalos are divided into five mass bins approximately spaced by $1\,\rm{dex}$, $[2.8\times10^9, \,2.3\times10^{10}, \,1.8\times10^{11}, \,1.4\times10^{12}, \,10^{13}, \,10^{14}]\,h^{-1}\,M_{\odot}$. For $M_{\rm{subhalo}} <10^{13}\,h^{-1}\,M_{\odot}$, the bin edges are set to 40 times the particle mass of each zoomed-in simulation, which is the minimum subhalo mass used when constructing the mock catalog. Correlation functions are computed using the Landy--Szalay estimator implemented in the CORRFUNC library \citep{Sinha2020}.

In Figure \ref{fig:Fig10_NASIM_subhalo_xi}, we show the two-point correlation functions $\xi(r)$ of subhalos measured in real space. The uncertainties in the correlation functions are estimated by partitioning the zoomed-in regions with a $(75\,h^{-1}\,\rm{Mpc})^3$ box to account for the variation relevant within the TNG100-1 box size. The mass-dependent clustering of subhalos agrees with TNG100-1-Dark across all mass bins; compared with TNG100-1, only one or two mass bins show slight differences on small scales. Considering the baryonic physics in TNG100-1 and its limited box size (e.g., only 139 subhalos in the most massive bin), we conclude that our zoomed-in simulations are consistent with TNG within the uncertainties.

\subsection{Application of MSSM to the $N$-body Simulation NASIM \label{sec:Sec4.3}}
We first prepare a subhalo catalog for each simulation box to which the MSSM will be applied. Subhalos with at least 40 particles are considered based on our simulation design. Under this minimum particle number, the mass function and clustering are in reasonable agreement with TNG, as shown in Figures \ref{fig:Fig9_NASIM_MassFunc} and \ref{fig:Fig10_NASIM_subhalo_xi}. All the input DM features are computed using the same definitions described in Section \ref{sec:Sec3.1}.

However, it is worth noting that most of the input features are resolution dependent. In particular, the quantities that are tightly coupled to the internal structure of halos show large discrepancies that arise from both resolution and intrinsic differences between DM-only and hydro simulations (see Appendix \ref{AppC}). Baryonic physics such as radiative cooling and feedback alters the internal structure of halos \citep[e.g.,][]{Duffy2010,DiCintio2014,Schaller2015,Anbajagane2022,Sorini2025}. Correcting for these effects is nontrivial even among hydrodynamical simulations due to differences in subgrid physics implementations. These halo properties also play a crucial role in MSSM (see Figure 7 of JK19 and Section \ref{sec:Sec5.1.1}), so that the predicted baryonic-property distributions closely follow those of the input halo properties.

\begin{figure}[t]
\centering
\includegraphics[width=0.99\linewidth]{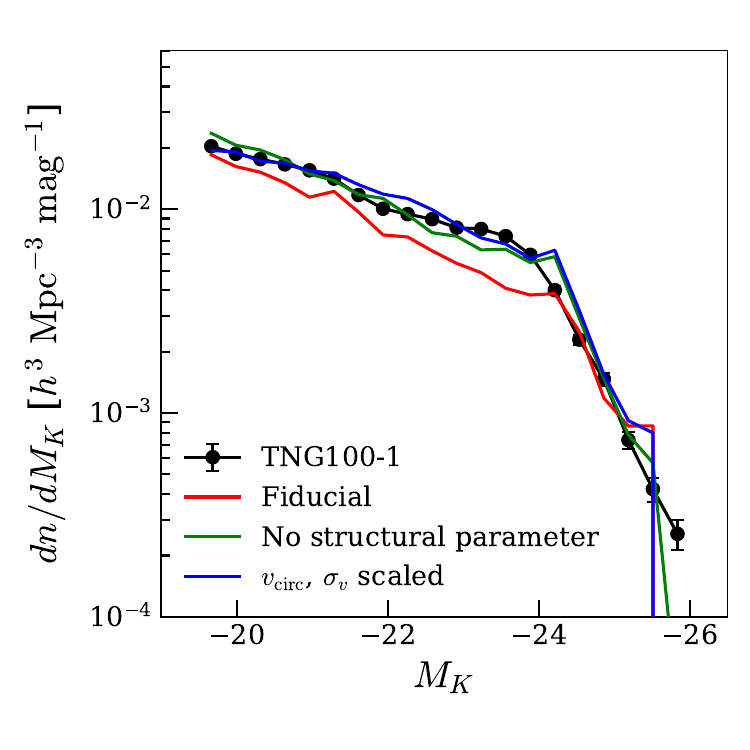}
\caption{$K$-band luminosity functions predicted with the $N$-body simulation when three different models are applied. The black line denotes the luminosity function of original TNG100-1 subhalos. Although the predictions are noisy, our two approaches are in reasonable agreement with TNG100-1.
\label{fig:Fig11_NASIM_K_LumFunc}}
\end{figure}

This mismatch between high-resolution hydrodynamical training data and lower-resolution $N$-body realizations is a generic challenge for any framework that transfers a learned halo--galaxy mapping across simulations of different resolution. Similar issues have been reported and addressed in the literature in various ways, for example by constructing a model variant that avoids an internal-structure feature \citep{Hausen2023} or by quantifying the impact of resolution and baryonic physics on the transferred model separately \citep{Chittenden2025}. In this context, we pursue two complementary approaches below, which together make the trade-off between accuracy and simulation independence explicit.

First, we train a model without structural parameters ($v_{\rm{circ}}$, $\sigma_v$, and subhalo anisotropy parameters), which are strongly simulation dependent. Although using the full set of input features yields the best performance (Section \ref{sec:Sec4.1}), the new parameters introduced in this study may partly compensate for the removal of the structural features. Even so, the model performance drops to $R^2=0.90,0.87,0.89,0.89,0.68,0.68$ ($\rm{MAE}=0.19,0.14,0.47,0.44,0.28,0.16$) for $M_{\rm{star}}$, $M_{\rm{gas}}$, $M_K$, $M_r$, SFR, and $Z_{\rm{gas}}$, respectively, when structural parameters are excluded. A practical advantage, however, is that this variant avoids recalibrating structural parameters between our $N$-body simulations and TNG100-1. Second, we scale $v_{\rm{circ}}$ and $\sigma_v$ in a mass-dependent manner to match TNG100-1. As shown in Appendix \ref{AppC}, the values in our $N$-body simulations depart from those of TNG100-1 only over a limited range (e.g., $v_{\rm{circ}}=150$--$350\,\rm{km}\,\rm{s}^{-1}$). Although this procedure is not entirely physical, it preserves the model performance. In Appendix \ref{AppC}, we present the scaling result for $v_{\rm{circ}}$ and $\sigma_v$; corrections of up to $20$\%--$30$\% are required. 
Figure \ref{fig:Fig11_NASIM_K_LumFunc} shows the three $K$-band luminosity functions predicted from the same subhalo catalog. Both approaches agree reasonably well with TNG100-1, in contrast to the fiducial model. However, when structural parameters are removed, the luminosity function departs more strongly--- especially at the faint end---and the overall performance decreases significantly, as noted above. We therefore adopt the model with mass-dependent scaling of $v_{\rm{circ}}$ and $\sigma_v$, scaling only these two parameters because they are the most influential for the predictions; in principle, the distributions of other structural quantities should also be matched.

In Appendix \ref{AppD}, we present the full application results of MSSM in a format identical to Figure \ref{fig:Fig7_Fiducial_Predict_Phi}. The trend is similar to that for the test set, namely baryonic properties related to stars---stellar mass, luminosity---show an overall agreement with TNG100-1 (note, however, that the ground truth of NASIM is unknown). However, the gas-related properties---gas mass, SFR, and metallicity---show differences at the tails of the distributions, consistent with the fact that the original model also failed to reproduce them.

While we adopt the second approach for our mock catalog, the first (structure-free) variant remains useful in contexts where its accuracy on relevant baryonic properties is sufficient (e.g., ${\sim}0.19\,\rm{dex}$ on $M_{\rm{star}}$, ${\sim}0.14\,\rm{dex}$ on $M_{\rm{gas}}$) and where avoiding the recalibration of structural parameters between the training and target simulations is preferable.

\subsection{Construction of Mock Data and Comparison to Observations \label{sec:Sec4.4}}

\begin{figure*}[p]
\centering
\includegraphics[width=0.99\textwidth]{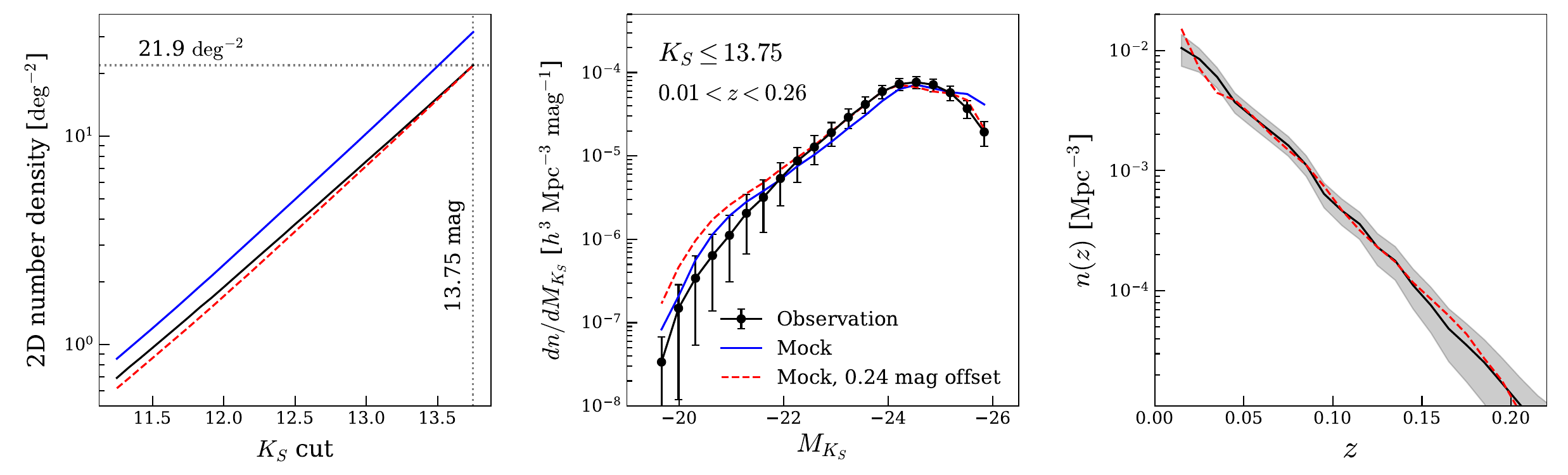}
\caption{Left: surface number density of target and mock galaxies as a function of apparent-magnitude cut. Middle: luminosity functions of galaxies with $K_S \leqslant 13.75$. Right: redshift distribution of galaxies. We add a $0.24\,\textrm{mag}$ offset to mock $M_{K_S}$ to match the surface number density of galaxies with $K_S \leqslant 13.75$. Measurements before and after applying the offset are shown as blue and red dashed lines. Except for the low-luminosity end, or equivalently low redshift, the target and mock galaxies are in reasonable agreement.
\label{fig:Fig12_Obs_Mock_LumFunc}}
\end{figure*}

\begin{figure*}[p]
\centering
\includegraphics[width=0.99\textwidth]{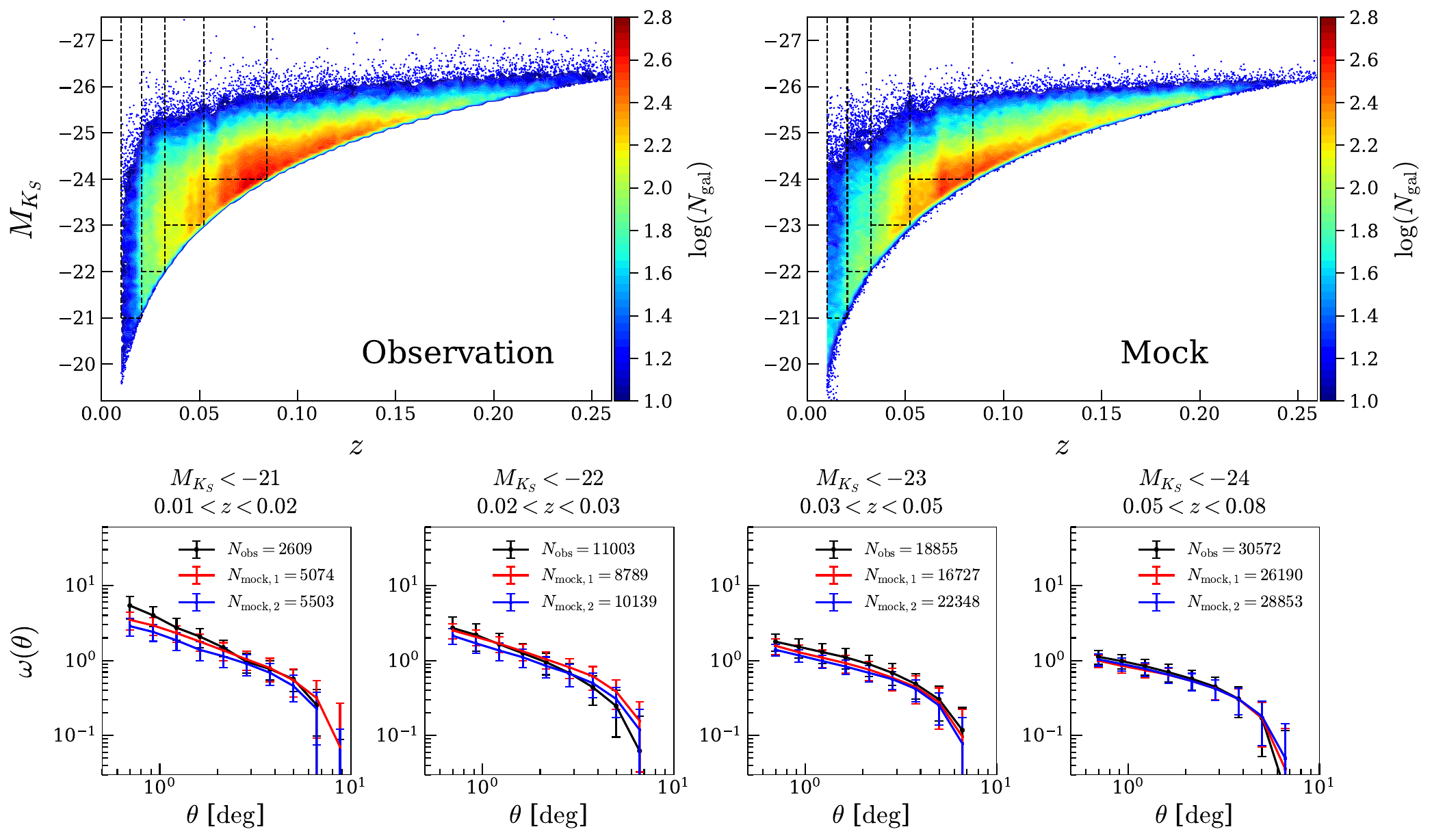}
\caption{Top: $M_{K_S}$ as a function of redshift for target and mock galaxies. The color map shows the number of galaxies ($N_{\rm{gal}}$). We show the galaxies residing in the region used for the clustering measurements. The mock galaxies are randomly removed according to the spectroscopic completeness. We compare the clustering of four volume-limited samples covering $M_{K_S}<[-24,-21]$ marked as black dashed lines. Bottom: angular two-point correlation functions $\omega(\theta)$ of four volume-limited samples defined in the top panels. Black lines are the measurements from the observation, and blue and red lines are from the mock data.
\label{fig:Fig13_Obs_Mock_z_MKs_omega}}
\end{figure*}

To construct mock data, we begin with the $M_K$ predicted in the previous section. Although a difference between $K$- and $K_S$-band magnitudes depends on the spectral energy distribution of galaxies, we add a small constant offset of $-0.044$ to all galaxies following \citet{Bessell2005}. Considering the accuracy of both observational data and our machine, this choice is not likely to affect the quality of the final mock data. We first compare the $K_S$-band luminosity function between predictions and observations. Because extrapolation is not possible with tree-based algorithms, predicted values are bounded by the physical range of the training dataset. This effect is pronounced at the high-luminosity end where there are more massive host subhalos in the $N$-body simulation (e.g., Figure \ref{fig:Fig11_NASIM_K_LumFunc}). Thus, we add $\sigma=0.3\,\textrm{mag}$ Gaussian scatter to the $K$-band magnitude when constructing the mock data. While the scatter is arbitrarily selected, we find that $\sigma\gtrsim0.3$ mag roughly reproduces the observed redshift distribution (see Appendix \ref{AppE} for details). In the left panel of Figure \ref{fig:Fig12_Obs_Mock_LumFunc}, we show the two-dimensional number density as a function of apparent magnitude cut. To account for the slight difference in the density, we add an offset of ${\sim}0.24\,\textrm{mag}$; this matches the density at $K_S \leqslant 13.75$. The middle panel shows the luminosity function of mock galaxies along with the one measured from the observation. As our mock data are also tailored to a flux-limited survey, we do not apply the $V_{\rm{max}}$ correction for apparent-magnitude cuts. The luminosity function from the observation is measured similarly to Equation \ref{eq:1}, but in a number of $M_{K_S}$ bins rather than redshift bins. There is a deviation around the low-luminosity end, but this is the regime where the local density significantly affects the measurement (i.e., cosmic variance). Thus, we conclude that our mock data are in reasonable agreement with the observation. A radial number density in redshift space after the correction is shown in the right panel.

After matching the number density, we compare the clustering of galaxies between the observation and the simulation. Two-point correlation functions of galaxies are simple and valuable tools for summarizing the distribution of galaxies. They are known to be dependent mainly on galaxy mass or luminosity, and also on the other properties such as color and specific SFR \citep{Li2006, Zehavi2011, Berti2021, Kakos2024}. Therefore, as a starting point, we compare the luminosity-dependent clustering of mock galaxies with that of the observation. Galaxies are divided into four volume-limited samples covering $M_{K_S}<[-24,-21]$ spaced by one magnitude. The boundaries of the volume-limited samples are shown as black dashed lines in the upper panels of Figure \ref{fig:Fig13_Obs_Mock_z_MKs_omega}. To minimize the spectroscopic incompleteness correction and the boundary effects due to discontinuity, we conservatively choose the region in the northern hemisphere with spectroscopic completeness larger than 0.5, amounting to $\approx19\%$ of the entire sky. For the random and mock datasets, we apply the same mask and remove galaxies randomly in each HEALPix pixel according to the spectroscopic completeness map constructed in Section \ref{sec:Sec2.1}. As correlation functions are only measured in a partial sky, we measure them with two datasets in the mock data by mirroring the mask to the southern hemisphere. The uncertainties are estimated by bootstrapping spatial blocks defined with $N_{\rm{side}}=4$ HEALPix \citep{Loh2008}. The bottom panels of Figure \ref{fig:Fig13_Obs_Mock_z_MKs_omega} show the angular correlation functions of four volume-limited samples. Except for the first luminosity bin where the volume is only $\approx2.7\times10^6\,\rm{Mpc}^3$, the number of galaxies in each volume-limited sample matches within $20\%$. The correlation functions show overall agreement with the observation. Although the first bin shows a slight difference on small scales, this sample is located at $z<$0.02, where the local density plays a crucial role. There is a noticeable difference in density between the mock data and the observation (see the rightmost panel of Figure \ref{fig:Fig12_Obs_Mock_LumFunc}). It has also been shown in \citet{Whitbourn2014} and \citet{Bohringer2020} that the Milky Way Galaxy is located in a local void of size ${\sim} 100\,\rm{Mpc}$, albeit with a shallower sample. While the discrepancy could be alleviated by choosing the observer location (center of the zoomed-in box in our case) that has a similar environment, we conclude that the luminosity-dependent clustering of our mock galaxy catalog is consistent with the observation.\\

\section{Discussion \label{sec:Sec5}}
In this section, we put forward several avenues for future research, mainly based on the characteristics of our machine, including feature importance and known caveats. In addition, we explore which environment measures are most important in our model and whether they aid in reproducing the gas mass fraction as a function of environment.

\subsection{Further Improvements in the Machine Learning Model \label{sec:Sec5.1}}

\subsubsection{Feature Importance of DM Features \label{sec:Sec5.1.1}}
Our ML model was able to reproduce the star-related quantities ($M_{\rm{star}}$, $M_K$, $M_r$) and $M_{\rm{gas}}$ accurately, but it showed limited ability to predict gas-related quantities (SFR, $Z_{\rm{gas}}$), especially for starburst and quiescent populations. This suggests that our adopted DM properties do not adequately encode the physics that regulates star formation and gas-phase metal enrichment. JK19 pointed out that MSSM is dictated by only a few input features such as $M_{\rm{subhalo}}$, $v_{\rm{circ}}$, and $\sigma_v$; it is essential to develop new important features to reproduce the diversity of galaxies. In light of this need, we adopted a number of features in this study for potential improvement. Figure \ref{fig:Fig8_MSSM_zoom3_summary} already shows that the input features slightly improve the model; here, we further investigate how much they contribute to the prediction of each baryonic property. We use the SHapley Additive exPlanations (SHAP) value \citep{Lundberg2017} to quantify the importance. The SHAP value, based on cooperative game theory, measures the contribution of each feature to prediction by considering all the possible combinations of features. We use \texttt{TreeExplainer} in the \texttt{SHAP} library to measure SHAP values with our machine.

\begin{figure}[t]
\centering
\includegraphics[width=0.99\linewidth]{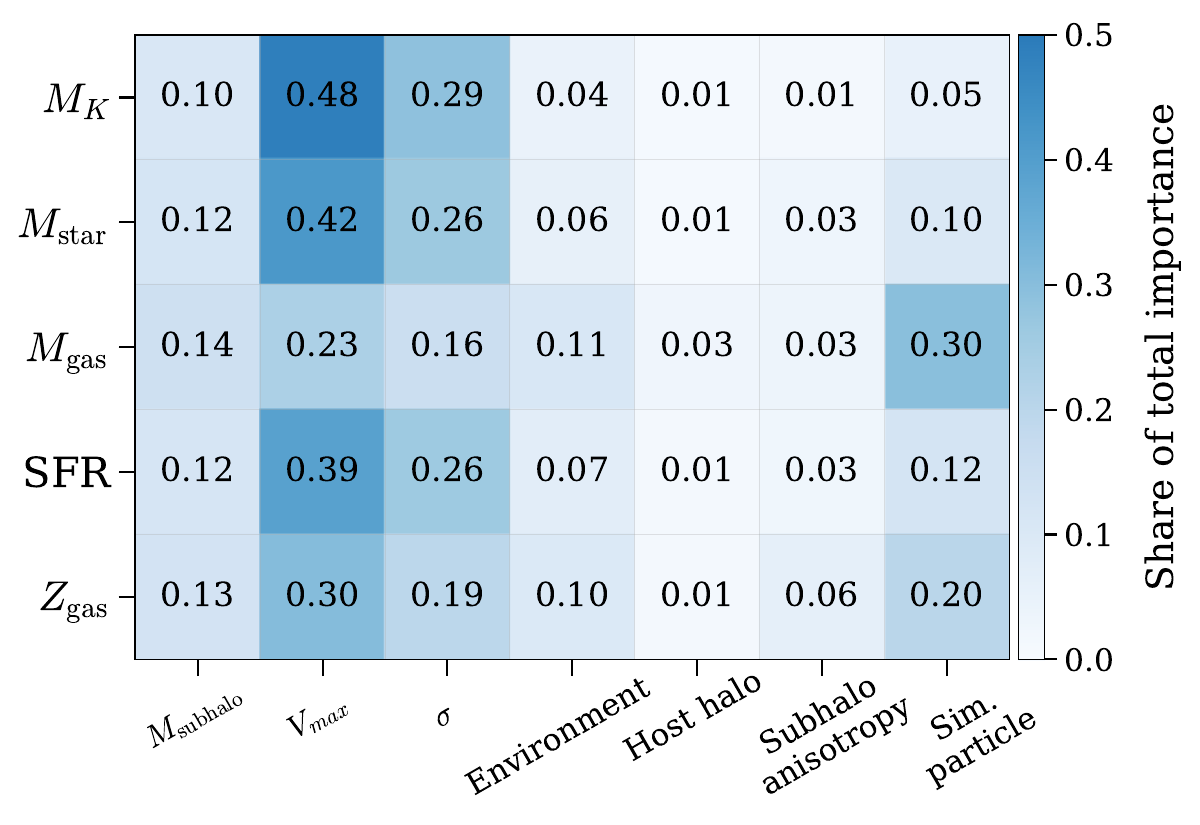}
\caption{Importance of feature groups defined by $\rm{mean}(|\rm{SHAP}~\rm{value}|)$ for each target baryonic property. The color map and numbers show the fraction of importance compared to the total importance. For stellar components, prediction is mostly dictated by $M_{\rm{subhalo}}$, $v_{\rm{circ}}$, and $\sigma_v$, but for gas-related properties, environment contributes more than $20\%$.
\label{fig:Fig14_Importance_Group}}
\end{figure}

Figure \ref{fig:Fig14_Importance_Group} shows the fraction of $\rm{mean}(|\rm{SHAP}~\rm{value}|)$ for five baryonic properties. As MSSM first predicts stellar magnitudes and utilizes them as an input, we add the contribution from this input when considering the second-stage target properties. The result is consistent with Figure \ref{fig:Fig8_MSSM_zoom3_summary}, where we added groups of new input features to the fiducial model and observed the improvement in predictive performance. The model is highly dependent on three key DM properties: $M_{\rm{subhalo}}$, $v_{\rm{circ}}$, and $\sigma_v$; they account for more than 50\% of the total importance for any baryonic property. In particular, star-related properties are well predicted solely based on these. A noteworthy pattern is that the gas-related properties (bottom three rows) depend strongly on other properties, accounting for $47$\%, $23$\%, and $37$\% of their total importance. These also showed improvement in Figure \ref{fig:Fig8_MSSM_zoom3_summary}, meaning that the adopted features actually assist in predicting those target properties. 
However, these gas-related properties are still not well predicted. Therefore, further investigation to identify important features for gas-related predictions or ML framework improvements is needed.\\

\subsubsection{Random Assignment of Unresolved Baryonic Properties \label{sec:Sec5.1.2}}

\begin{figure*}[t]
\centering
\includegraphics[width=\textwidth]{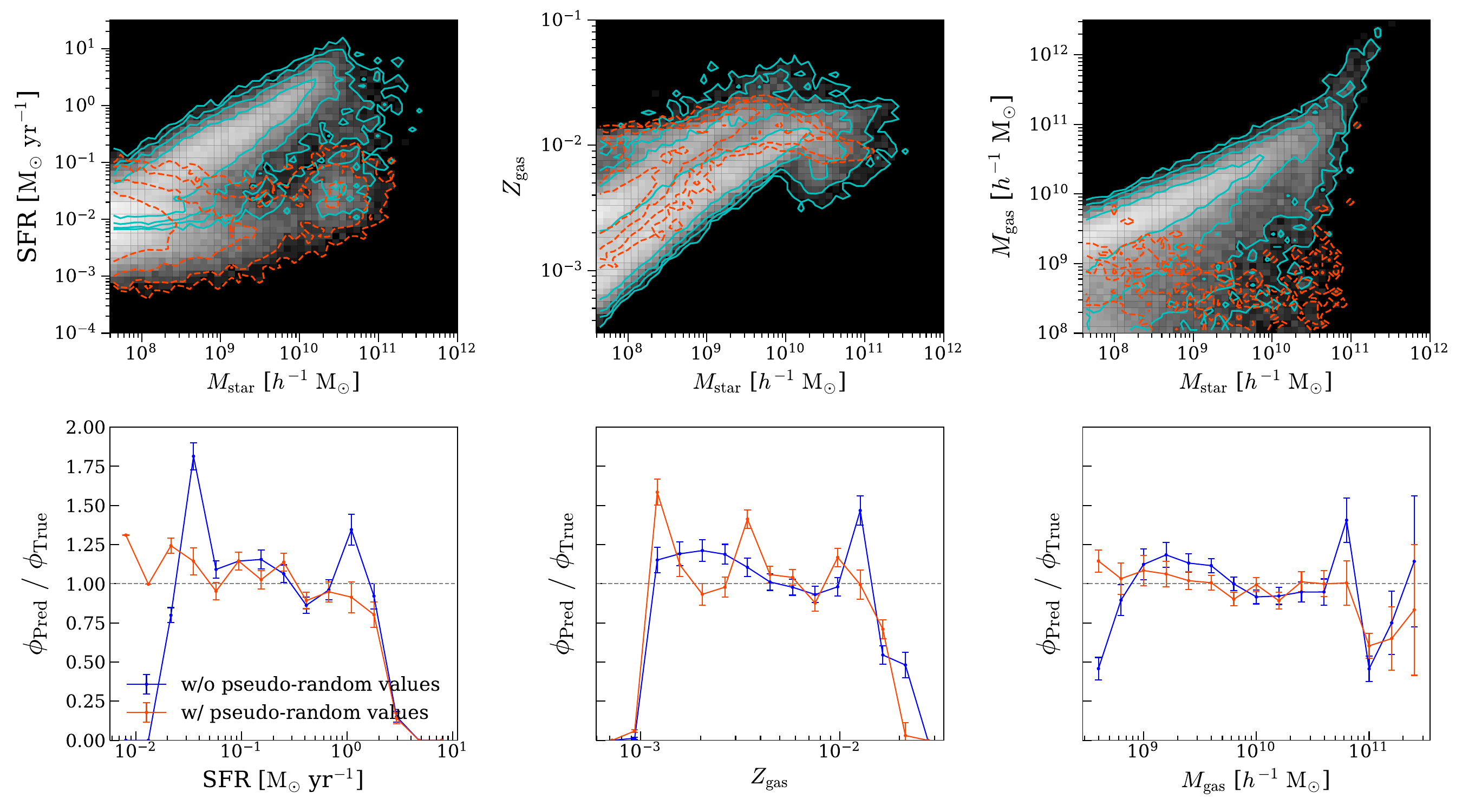}
\caption{Top: distribution of baryonic properties after random assignment as a function of stellar mass. SFR, $Z_{\rm{gas}}$, and $M_{\rm{gas}}$ are shown from the left to right panels. The gray color map in the background shows the entire population of subhalos, and the cyan (orange) contour shows the distribution of the resolved (unresolved) population in the original catalog. Bottom: the ratio between predicted and true distribution functions $\phi_{\rm{Pred}}$/$\phi_{\rm{True}}$ of gas-related baryonic properties, which is identical to the bottom panels of Figure \ref{fig:Fig7_Fiducial_Predict_Phi}. The random assignment of unresolved baryonic properties qualitatively improves the model, yielding a more accurate recovery of the original distribution function except for $Z_{\rm{gas}}$.
\label{fig:Fig15_unresolved_extrapolate}}
\end{figure*}

In an ERT, if a feature is truncated or highly skewed, informative thresholds concentrate within a narrow range. The model may form unbalanced partitions, resulting in systematic bias, especially near the boundary. For example, quiescent and low-mass galaxies exhibit intrinsically low SFRs and gas contents, which are often unresolved due to the resolution limits of simulations. In our study, nearly 40\% of the subhalos selected based on DM and stellar contents are excluded when applying the ($M_{\rm{gas}}$, SFR, $Z_{\rm{gas}}$) cuts defined earlier in Section \ref{sec:Sec3.1}. Modeling these unresolved galaxy populations remains challenging even in higher-resolution simulations, as observational constraints are also incomplete due to instrumental sensitivity limits (e.g., low-H$\alpha$ flux). However, we test whether assigning pseudorandom values to unresolved properties can mitigate biases. Specifically, we utilize the scaling relations between various indicators and gas-related properties ($M_{\rm{gas}}$, SFR, $Z_{\rm{gas}}$). The proxies may be any simulation-derived quantities tied to the stellar components of subhalos (e.g., colors, stellar-to-halo mass ratio), as each subhalo is constructed to contain at least 40 stellar particles. Such relations between proxies and baryonic properties have been observed for massive galaxies \citep{Kannappan2004, Tremonti2004, Salim2007, Wyder2007, Catinella2010}, albeit with scatter that depends on the mass range. Although these relations may not perfectly hold for low-mass galaxies, we adopt this approximate painting approach because our primary goal is to roughly populate the unresolved regime. 

We assign baryonic properties based on the radial basis function (RBF) interpolation implemented in the \texttt{Scipy} library. Missing values are interpolated or extrapolated based on the distance to the neighboring subhalos in multidimensional indicator space. This method is analogous to the color--color diagram analysis, where the galaxies are roughly divided into multiple populations depending on the location in a certain color space. We use four indicators ($\log M_{\rm{star}}$, $M_U-M_V$, $M_V-M_K$, $M_U-M_g$) of the 64 nearest neighbors and estimate missing values with a cubic kernel and linear extrapolation. Each indicator is normalized to be in the range $[0,1]$ to compensate for the difference in dynamic range. We do not pursue further optimization of the kernel or hyperparameters, as the potential benefit of a more precise treatment is expected to be limited within our current modeling framework. We add $0.3\,\rm{dex}$ scatter to the SFR and gas mass, where most missing values lie in a regime with limited knowledge from the resolved population.

In the upper panels of Figure \ref{fig:Fig15_unresolved_extrapolate}, we show the distribution of gas-related baryonic properties as a function of stellar mass. Gray color maps in the background show the distribution of the entire set of subhalos, and cyan and orange contours show the distribution of the original sample (resolved) and the filled sample (unresolved), respectively. We train our machine again with the subhalo catalog filled with random baryonic properties. The lower panels of Figure \ref{fig:Fig15_unresolved_extrapolate} show the ratios between true and predicted distribution functions before and after the pseudorandom value assignment. SFR and gas mass show qualitative improvement especially at the lower tails, although gas metallicity does not improve much. Based on our experiment, we suggest that a more precise treatment of the unresolved population (e.g., modeling spectra) has the potential to improve the machine performance.\\

\subsubsection{Additional Considerations and Caveats \label{sec:Sec5.1.3}}
In the previous sections, we discussed two aspects where the machine can be improved. However, there are several points and caveats that need further discussion and investigation. Below, we list items that may be important in future work.
\begin{enumerate} 
    \item As mentioned in Section \ref{sec:Sec3.1}, our model did not include the merger history of subhalos. Instead, we included information on a fixed timescale $\Delta t_{z=0.2}\sim2.5\,\rm{Gyr}$ using the approximate method. Using realistic mass flow from a merger tree will increase the predictability of baryonic properties such as the SFR that are tightly coupled to it. Moreover, mass flow measured with different timescales would also be helpful, as the dynamical timescale is different for each subhalo. In addition to the mass accretion rate, the merger history itself could provide valuable information, as recent mergers significantly affect the physical properties of both subhalos and galaxies.

    \item\label{Subhalo_Property} Subhalo properties are different depending on the type (e.g., baryonic physics, resolution) of simulation. As discussed in Appendix \ref{AppC}, there is a discrepancy in subhalo structure between TNG100-1 and our $N$-body simulations, as well as in TNG100-1-Dark, which has a resolution comparable with TNG100-1. While we have scaled the key input features ($v_{\rm{circ}}$, $\sigma_v$) to mitigate the issue, a more precise treatment is required, as the scaling factor would depend on more than the mass. This is also the case for TNG300-1, where we used a constant scaling factor to overcome the resolution difference. If we can properly scale subhalo properties between different resolution runs, less compromise between resolution and volume will be required (e.g., combining knowledge from all TNG runs). For example, an ML model such as that of \citet{Jung2024} could be one of the potential avenues for future studies.

    \item Alongside \ref{Subhalo_Property}, subhalo populations are intrinsically different between hydrodynamical and DM-only simulations due to baryonic physics complications (e.g., the destruction of subhalos). We excluded subhalos with too few DM and stellar particles owing to resolution limitations and because they fall outside the mass range relevant to this study. However, there are small populations of physical subhalos with very low contents of stars or DM \citep{Benitez2017, Haslbauer2019, Shin2020, Lee2024}. As the formation of these subhalos is tightly related to the environment and history, they should also be considered by the machine. This will become more important in the future as the predictive performance of the machine improves (as well as the resolution of the hydrodynamical simulations).\\
\end{enumerate}

\subsection{How Effective Are Our Environmental Features? \label{sec:Sec5.2}}

The role of environment in the evolution of galaxies has long been investigated using various measures. While it is well established that more massive and redder galaxies are more likely to reside in dense regions, there remains no universal metric that can encompass the full complexity of the galaxy--environment connection. One widely used approach is to quantify the local density measured within a circular aperture---either within a fixed radius or based on the distance to the nearest neighbors---which has proven effective in distinguishing galaxy populations \citep[see][and references therein]{Muldrew2012}. Moreover, recent studies have suggested that density alone is not sufficient to capture all environmental effects and that additional factors, such as the traceless components of the tidal tensor, become important. The relative positions of galaxies within the large-scale structure---such as nodes, filaments, sheets, and voids---play an additional role even at fixed mass and density \citep{Kraljic2018, Laigle2018, Sarron2019, Bonjean2020, Song2021}. Nevertheless, a clear consensus has not yet been reached on the most effective definition of the galaxy–environment connection, and disentangling genuine environmental effects from the dominant influence of DM halos remains a nontrivial task.

\begin{figure}[t]
\centering
\includegraphics[width=0.99\linewidth]{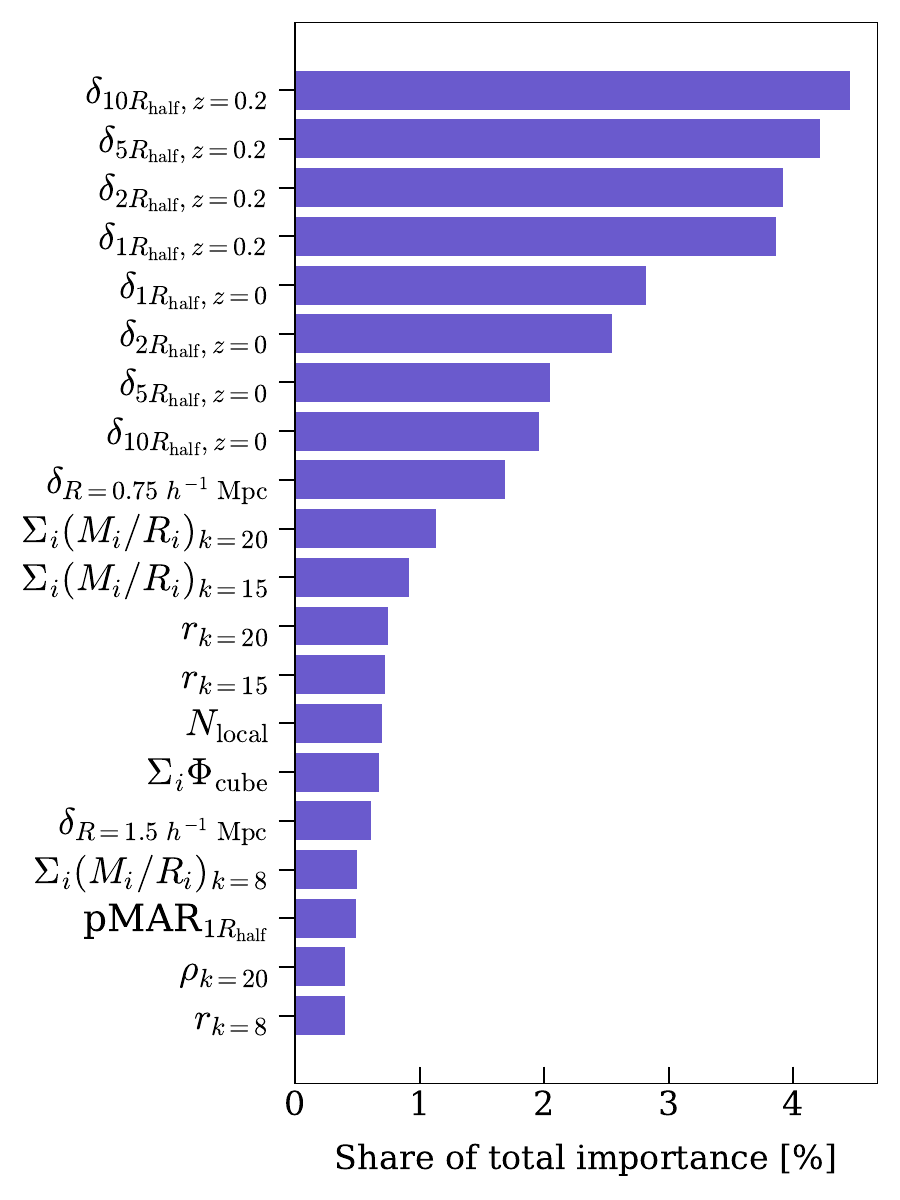}
\caption{The same as Figure \ref{fig:Fig14_Importance_Group}, but for $M_{\rm{gas}}$ decomposed into the top 20 important environment measures adopted in this study. $\delta$ measured in a window proportional to the size of subhalo contributes the most, particularly at $z=0.2$. $\delta$ measured at the fixed scale of $0.75\,h^{-1}\,\rm{Mpc}$ and (potential, distance) measured with the $k=15,20$ neighboring subhalos are also important in predicting $M_{\rm{gas}}$.
\label{fig:Fig16_Importance_GasMass_Env}}
\end{figure}

\begin{figure*}[t]
\centering
\includegraphics[width=0.9\textwidth]{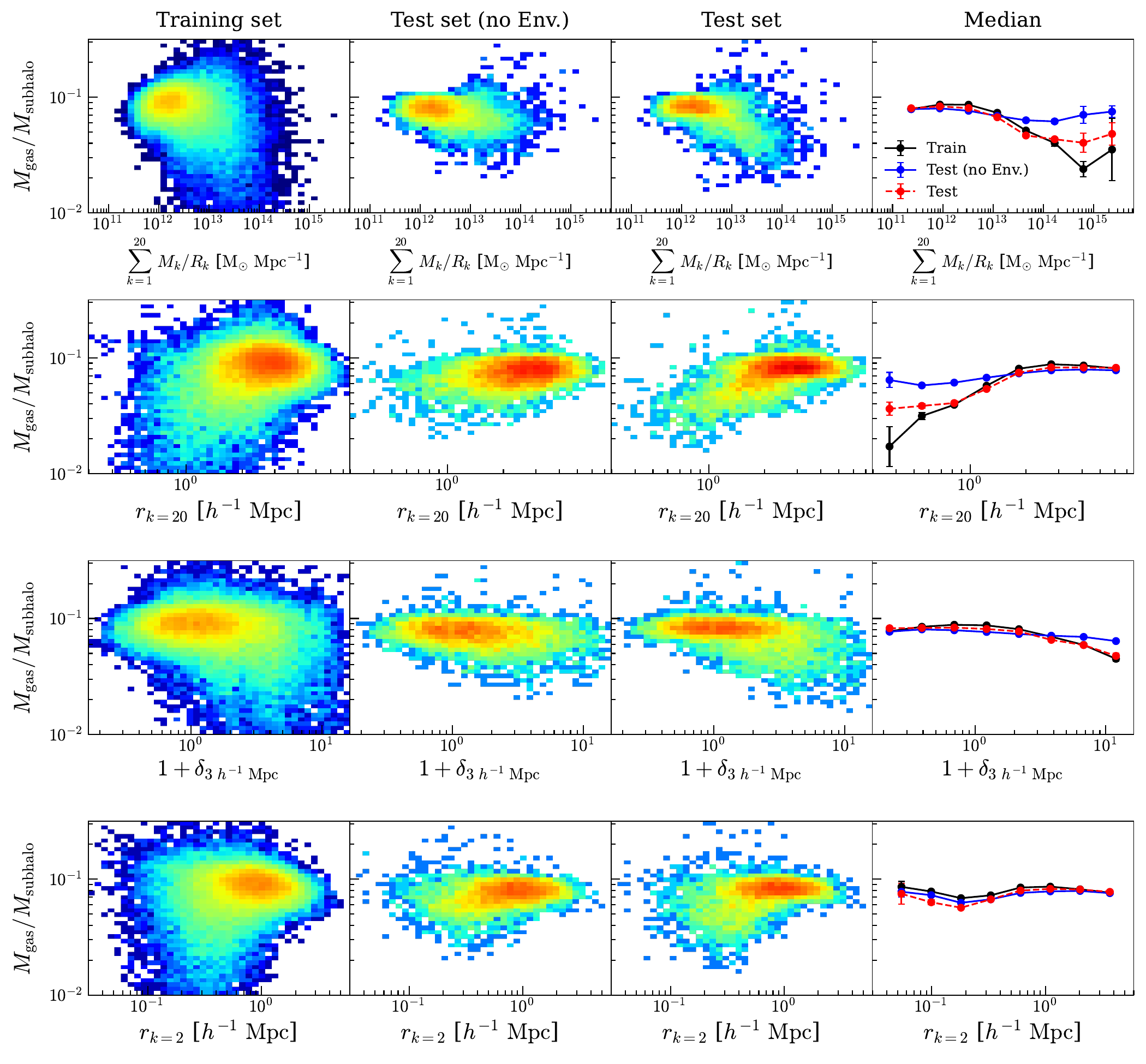}
\caption{Distribution of gas mass fraction $M_{\rm{gas}}$/$M_{\rm{subhalo}}$ as a function of four different environment definitions. The first column shows the distribution of the training dataset, and the second (third) column shows the distribution of predicted values when environmental features are excluded (included) from the input. The rightmost column shows the binned values as a function of environment. $M_{\rm{gas}}$/$M_{\rm{subhalo}}$ varies more significantly when using the higher SHAP value features, and environment--$(M_{\rm{gas}}$/$M_{\rm{subhalo}})$ relations are recovered only when environmental features are used in the ML model. 
\label{fig:Fig17_GasFrac_Env}}
\end{figure*}

The environment has been studied with ML in \citet{Xu2021}, \citet{Delgado2022}, \citet{dS2022}, \citet{Lovell2022}, \citet{Hernandez2023}, \citet{Wu2024}, \citet{Xu2024}, and \citet{Uchida2025}, focusing on feature importance and improvement in model accuracy. Following previous studies, we investigate which definition of environment affects our model the most and whether our machine reproduces the environmental scaling relations, with our analysis focusing mainly on $M_{\rm{gas}}$. In Section \ref{sec:Sec3.1}, we defined environments in various ways (seven definitions; 52 parameters in total across scales) including the ones widely adopted in observations. We first decompose the total importance for predicting $M_{\rm{gas}}$ (calculated in Section \ref{sec:Sec5.1.1}; the third row of Figure \ref{fig:Fig14_Importance_Group}) into individual environment measures. In this analysis, we again focus on the subhalos corresponding to the resolution of the zoom3 simulation ($M_{\rm{subhalo}}>2.3\times10^{10}\,h^{-1}\,M_{\odot}$), where the median redshift of the target galaxies lies. The mean number density of subhalos is approximately $0.12\,h^{3}\,\rm{Mpc}^{-3}$. Figure \ref{fig:Fig16_Importance_GasMass_Env} shows the top 20 highest SHAP value environmental parameters defined either with neighboring subhalos or simulation particles. The highest contributions clearly come from $\delta$ measured in a window proportional to $R_{\rm{half}}$. In particular, $\delta$ measured at $z=0.2$ is more important than that measured at $z=0$. Although we did not utilize the full merger tree, this may reflect the effect of tidal stripping on DM subhalos (e.g., the subhalo properties before infall describe the galaxy properties better than those at the current epoch). This result also suggests that a window proportional to the size (or mass) of a subhalo is more influential than one of fixed scale. However, we note that the scale probed by the half-mass radius is relatively small compared to the fixed scale adopted in this study, especially for less massive subhalos (e.g., median $10 \times R_{\rm{half}}\approx[0.3, 0.6, 1.7, 3.5]\,h^{-1}\,\rm{Mpc}$ for $[10^{11}, 10^{12}, 10^{13}, 10^{14}]\,h^{-1}\,M_{\odot}$ subhalos in our TNG100-1 sample); further assessment is required. For the $\delta$ smoothed with a fixed scale, the contribution from the smallest scale $R=0.75\,h^{-1}\,\rm{Mpc}$ is the highest and decreases with scale. Anisotropic environment measures $\alpha_R$ contributed the least (not shown in the figure), contributing less than 1\% in total. Conversely, environments measured with neighboring subhalos are effective especially in the case of potential ($\Sigma_{i} (M_i/R_i)$) and distance ($r_k$) measured with $k=15,20$ neighbors.

Because SHAP values only show how much predictions vary with the input features, they do not necessarily guarantee a physical connection between environment and galaxy properties. To confirm that our machine is able to reproduce the environment--galaxy relation, we investigate the gas mass fraction ($M_{\rm{gas}}/M_{\rm{subhalo}}$) as a function of various environment measures. The gas mass fraction is expected to decrease as the density increases at the current epoch. While a specific SFR and $Z_{\rm{gas}}$ might also show similar trends, we focus on the gas mass considering the machine accuracy. We choose four quantities, $\Sigma_{k=20} (M_k/R_k)$, $r_{k=20}$, $(1+\delta_{3\,h^{-1}\,\rm{Mpc}})$, and $r_{k=2}$, with a wide range of importance based on Figure \ref{fig:Fig16_Importance_GasMass_Env}. Two quantities measured using $20$ neighbors are among the most important features, and $r_{k=2}$ has a low contribution to the prediction. We add the simple overdensity $(1+\delta_{3\,h^{-1}\,\rm{Mpc}})$, which has modest importance, for comparison. This test is also a cross validation of whether the high-SHAP value features (e.g., $\Sigma_{k=20} (M_k/R_k)$) are more informative environment measures than the low-SHAP value features (e.g., $r_{k=2}$). Figure \ref{fig:Fig17_GasFrac_Env} shows the gas mass fraction as a function of the environment measures selected above. The first column shows the distribution of our training set, and the second and third columns show the result for the test dataset, but without and with environmental features either computed with neighbors or simulation particles. The rightmost column shows the median values measured in eight bins where the errors are measured by 1000 bootstrap resamples. The gas mass fraction clearly decreases in overdense environments, and the decrement is particularly significant when using $\Sigma_{k=20} (M_k/R_k)$ and $r_{k=20}$. This is consistent with Figure \ref{fig:Fig16_Importance_GasMass_Env}, where the two measures had larger importance than the other two. Nonetheless, $(1+\delta_{3\,h^{-1}\,\rm{Mpc}})$ also shows a clear trend even though it is smaller than the previous cases. Conversely, $r_{k=2}$ does not show a clear trend even in the original data. While the nearest subhalos may contain valuable information, as the matter can be directly transferred, our machine was not able to fully utilize this information. One important trend is the difference between models with (red dashed lines) and without (blue solid lines) environmental features. The model with the environmental parameters is able to reproduce the original environment--$(M_{\rm{gas}}/M_{\rm{subhalo}})$ relation. In contrast, the other model fails and shows an almost constant gas fraction in all cases. Accordingly, we conclude that our environmental features are indeed effective in capturing the impact of environment on gas fraction in galaxies. It is also possible to sort out the most relevant environmental features using the ranking of SHAP values. However, we have not made a full comparison of all the environmental features used in this study, and interpreting high SHAP values as the effectiveness of an environment measure should be assessed carefully. 

Although the distances to the neighboring subhalos (or galaxies) depend on the sample density, information about neighboring subhalos---especially potential and distance---has been shown to be informative. Therefore, incorporating additional properties of neighboring subhalos, such as their velocities, spatial distribution (rather than simple distance), and physical properties, would be a promising avenue to improve the machine's performance.

\section{Summary \label{sec:Sec6}}
We have used MSSM to connect DM and baryon components of subhalos in the IllustrisTNG simulation. With this halo-painting machine, we constructed a mock galaxy catalog for A-SPEC from an $N$-body simulation that matches the number density and clustering of the target galaxies. The principal results of this study are as follows.

\begin{enumerate}
    \item We trained MSSM using the subhalos with 40 or more particles for both DM and stellar components in TNG100-1. The model has the capacity to predict baryonic properties with $R^2=0.959$, $0.895$, $0.701$, and $0.792$ for $M_{\rm{star}}$, $M_{\rm{gas}}$, SFR, and $Z_{\rm{gas}}$, respectively. We adopted additional input features: (1) subhalo anisotropy parameters, (2) environment based on the nearest neighbors, and (3) simulation particles to improve the predictive performance. Adopting these features slightly improved the accuracy for stellar components and substantially improved it for gas metallicity. However, the predictive performance for gas-related properties, particularly SFR and $Z_{\rm{gas}}$, is still low, which requires more studies to find important input features or machine architectures that properly capture the star formation activity and metal enrichment of galaxies.

    \item We investigated the feature importance of our model based on SHAP values. $M_{\rm{subhalo}}$, $v_{\rm{circ}}$, and $\sigma_v$ account for more than 50\% of the total importance in any case of baryonic properties. However, the importance of environment increases to $20$\%--$40$\% in the case of gas-related properties. We decomposed the importance into individual environment measures and found that density contrast $\delta$ measured at $z=0.2$ within a window proportional to subhalo size affected our prediction the most. Among the environmental parameters measured with the nearest neighbors, $k=15,20$ had higher importance than others for a $0.12\,h^{3}\,\rm{Mpc}^{-3}$ number density sample. Our analysis suggests that including environmental features aids the machine in recovering the environment--$(M_{\rm{gas}}/M_{\rm{subhalo}})$ relation.

    \item We conducted four zoomed-in $N$-body simulations tailored to match the requirements of A-SPEC. The zoomed-in regions have a progressively higher resolution near the center of the simulation box, mimicking the flux-limited survey. We applied MSSM to the subhalo catalogs from the zoomed-in simulations to construct a mock galaxy catalog. The mock galaxy catalog reproduces the luminosity function and luminosity-dependent clustering of target galaxies when tuned to match the number density.

\end{enumerate}

\noindent Our mock galaxy catalog delivers reliable stellar masses, stellar luminosities, and gas masses, with coarse estimates of the SFR and gas metallicity. The catalog\footnote{The data are available on Zenodo at \dataset[doi:10.5281/zenodo.21140927]{https://doi.org/10.5281/zenodo.21140927}.} will serve as a foundational resource for the planned science program of A-SPEC, including the construction of a complete catalog of galaxy groups and clusters in the nearby Universe (K. Dachan et al. 2026, in preparation).

\begin{acknowledgments}
We thank the referee for constructive comments.
D.K., H.S.H., and J.-h.K. are supported by the Global-LAMP Program of the National Research Foundation of Korea (NRF) grant funded by the Ministry of Education (No. RS-2023-00301976).
H.S.H. also acknowledges support from the National Research Foundation of Korea (NRF) grant funded by the Korea government (MSIT; RS-2026-25482692).
This work was partially supported by the Korea Astronomy and Space Science Institute under the R\&D program (Project No. 2026-1-831-03), supervised by the Korea Aerospace Administration.
J.-h.K.'s work was supported by the National Research Foundation of Korea (NRF) grant funded by the Korea government (MSIT) (Nos. 2022M3K3A1093827 and 2023R1A2C1003244). His work was also supported by the National Institute of Supercomputing and the Network/Korea Institute of Science and Technology Information with supercomputing resources including technical support, grants KSC-2021-CRE-0442, KSC-2022-CRE-0355, and KSC-2024-CRE-0232. H.S. was supported by the National Research Foundation of Korea (NRF) grant funded by the Korea government (MSIT) (No. 2022M3K3A1093827). M.Y. is supported by the Korea Astronomy and Space Science Institute grant funded by the Korea government (KASA, Korea AeroSpace Administration) (2026183201). The work of H.B. was supported by the Basic Science Research Program through the National Research Foundation of Korea (NRF) funded by the Ministry of Education (RS-2025-25403440). C.P. is supported by the KIAS Individual grant PG016904 at the Korea Institute for Advanced Study (KIAS) and by the National Research Foundation of Korea (NRF) grant funded by the Korean government (MSIT; RS-2024-00360385).

This work is conducted as part of the A-SPEC program, which utilizes the K-SPEC instrument mounted on the KMTNet 1.6 m telescope operated by the Korea Astronomy and Space Science Institute (KASI) at Siding Spring Observatory (SSO), Australia. Data transfer infrastructure between SSO and KASI is supported by the Korea Research Environment Open NETwork (KREONET).
A-SPEC is an all-sky spectroscopic survey of nearby galaxies conducted with the K-SPEC instrument. A-SPEC is managed by the Korean Spectroscopic Survey Consortium for the Participating Institutions including the Korea Astronomy and Space Science Institute (KASI), the Korea Institute for Advanced Study (KIAS), and Seoul National University (SNU).

This work is also
supported by the Center for Advanced Computation at the Korea
Institute for Advanced Study.
The analysis presented in this work was partially performed on the Olaf supercomputer, operated by the Research Solution Center at the Institute for Basic Science.
\end{acknowledgments}

\software{
  Numpy \citep{Numpy2020},
  Scipy \citep{Scipy2020},
  matplotlib \citep{Matplotlib},
  astropy \citep{2013A&A...558A..33A,2018AJ....156..123A,2022ApJ...935..167A},
  healpy \citep{Healpy2019,Gorski2005},
  scikit-learn \citep{scikit-learn}
}

\appendix

\section{Scaling TNG300-1 Baryonic Properties to TNG100-1}
\label{AppA}
\begin{figure*}[t]
\centering
\includegraphics[width=0.7\textwidth]{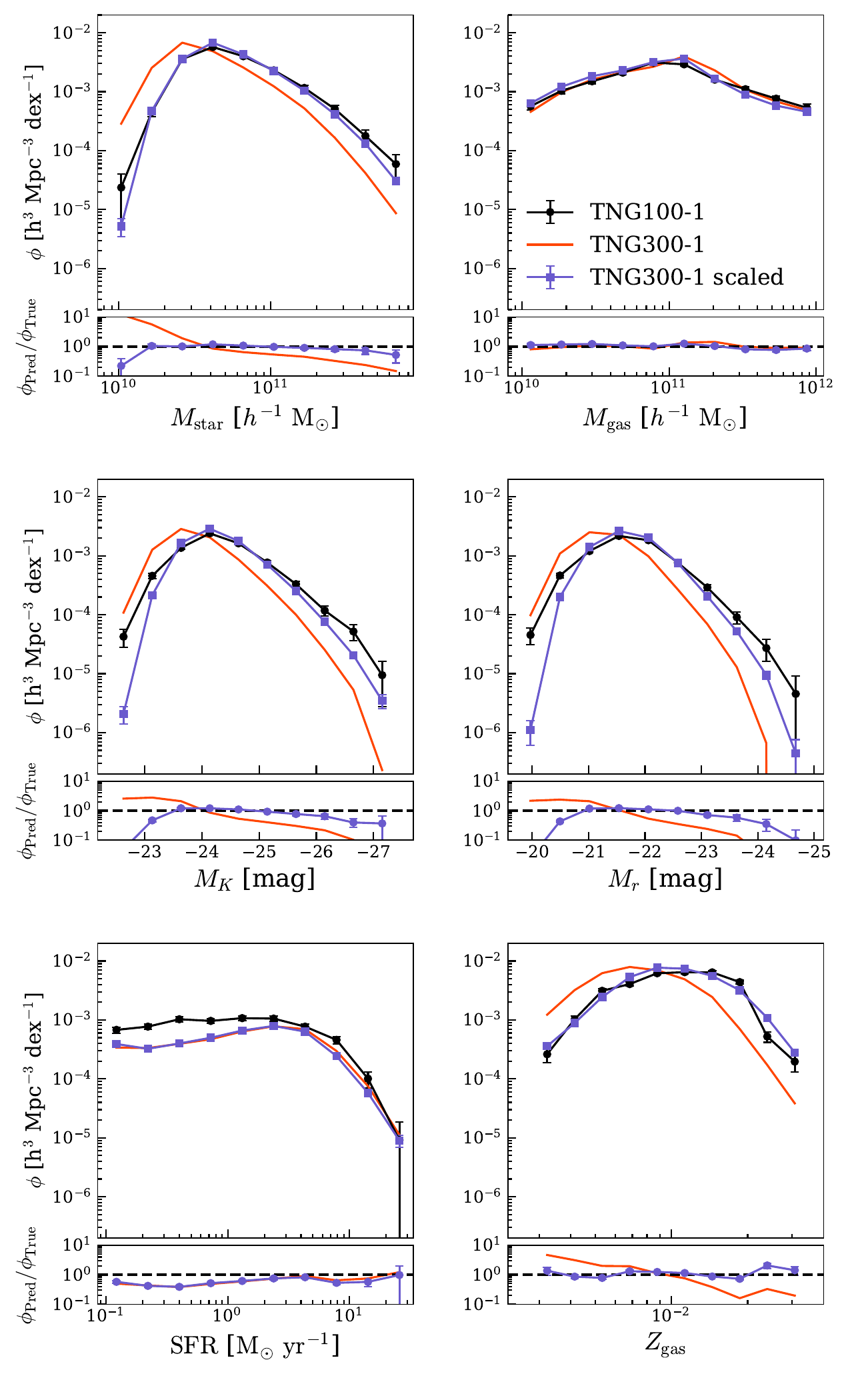}
\caption{Distribution functions $\phi$ of six key baryonic properties for TNG100-1 and TNG300-1 subhalos more massive than $10^{12}\,h^{-1}\,M_{\odot}$. The orange and blue lines show the $\phi$ before and after the scaling factors are applied.  
\label{fig:AppendixA_TNG300_TNG100_scale_distribution}}
\end{figure*}

In this appendix, we describe the scaling factor adopted in Section \ref{sec:Sec2.2} to merge the TNG300-1 subhalo catalog into that of TNG100-1. We first choose subhalos more massive than $10^{12}\,h^{-1}\,M_{\odot}$ from the two simulations. We keep the subhalo mass limit as low as possible to encompass a wide range of baryonic properties, even though the scaling factor might not be constant over the mass range we selected. Moreover, if we restrict our samples to more massive halos, the statistics of the reference catalog (TNG100-1) will be intrinsically noisy due to the low number of samples. With this sample in hand, we investigate the scaling factors in the range $[0.5, 2]$ and find the scaling factor that minimizes the difference in the distribution functions. We show the distribution functions of key baryonic properties before and after the scaling in Figure \ref{fig:AppendixA_TNG300_TNG100_scale_distribution}, and each scaling factor that minimizes the discrepancy is tabulated in Table \ref{tab:TabA1}. The stellar mass and luminosities are consistent with the scaling factor suggested in \citet{Pillepich2018}, and gas-related properties show different values depending on the properties. We note, however, that the scaling is not perfect, as shown in Figure \ref{fig:AppendixA_TNG300_TNG100_scale_distribution}, and strategies to increase the training sample through TNG300-1 should be further investigated in future work.

\begin{deluxetable*}{cc}
\tablecaption{Scaling Factors Used for TNG300-1 Subhalos \label{tab:TabA1}}
\tablewidth{0pt}

\tablehead{
\colhead{Baryonic Property} & \colhead{Scaling Factor}
}

\startdata
$M_{\rm{star}}$ & 1.44 \\
$M_{\rm{gas}}$  & 0.83 \\
$M_U$           & 1.43 \\
$M_B$           & 1.47 \\
$M_V$           & 1.45 \\
$M_K$           & 1.45 \\
$M_g$           & 1.45 \\
$M_r$           & 1.47 \\
$M_i$           & 1.45 \\
$M_z$           & 1.44 \\
SFR             & 1.19 \\
$Z_{\rm{gas}}$  & 1.39 \\
\enddata
\end{deluxetable*}

\section{Full Metric Table of Models with Different Input Features}
\label{AppB}
In this appendix, we present the full metric table (Table \ref{tab:AllModelPerf}), which reports the performance of each model described in the main text.

\begin{deluxetable*}{lcccccccccccc}
\tabletypesize{\scriptsize}
\renewcommand{\arraystretch}{1.05}
\tablecaption{Predictive Performance of All Models Considered in This Study\label{tab:AllModelPerf}}
\tablewidth{0pt}
\tablehead{
\multicolumn{5}{c}{} & \multicolumn{4}{c}{\textit{Part 1 of 2}} & \multicolumn{4}{c}{} \\
\cline{2-5}\cline{6-9}\cline{10-13}
 & \multicolumn{4}{c}{Fiducial} & \multicolumn{4}{c}{kNN Aperture} & \multicolumn{4}{c}{Host Halo} \\
\colhead{Output} & \colhead{$R^2$} & \colhead{$\Delta R^2$} & \colhead{MAE} & \colhead{$\rho$} & \colhead{$R^2$} & \colhead{$\Delta R^2$} & \colhead{MAE} & \colhead{$\rho$} & \colhead{$R^2$} & \colhead{$\Delta R^2$} & \colhead{MAE} & \colhead{$\rho$}
}
\startdata
$M_{\rm{star}}$ & 0.955(0.001) & 0.006 & 0.123 & 0.977 & 0.954(0.001) & 0.006 & 0.125 & 0.977 & 0.955(0.001) & 0.006 & 0.123 & 0.977 \\
$M_{\rm{gas}}$  & 0.887(0.004) & 0.007 & 0.132 & 0.943 & 0.888(0.003) & 0.010 & 0.130 & 0.943 & 0.886(0.004) & 0.010 & 0.132 & 0.942 \\
$M_{U}$         & 0.846(0.003) & 0.010 & 0.516 & 0.920 & 0.843(0.003) & 0.004 & 0.520 & 0.918 & 0.846(0.003) & 0.006 & 0.515 & 0.920 \\
$M_{B}$         & 0.890(0.003) & 0.008 & 0.433 & 0.943 & 0.888(0.002) & 0.003 & 0.436 & 0.942 & 0.890(0.002) & 0.005 & 0.432 & 0.944 \\
$M_{V}$         & 0.915(0.002) & 0.006 & 0.384 & 0.957 & 0.914(0.002) & 0.003 & 0.388 & 0.956 & 0.916(0.002) & 0.004 & 0.383 & 0.957 \\
$M_{K}$         & 0.937(0.001) & 0.005 & 0.361 & 0.968 & 0.936(0.001) & 0.003 & 0.365 & 0.967 & 0.937(0.001) & 0.003 & 0.361 & 0.968 \\
$M_{g}$         & 0.900(0.002) & 0.007 & 0.414 & 0.949 & 0.898(0.002) & 0.003 & 0.417 & 0.948 & 0.900(0.002) & 0.004 & 0.413 & 0.949 \\
$M_{r}$         & 0.922(0.002) & 0.006 & 0.370 & 0.960 & 0.921(0.002) & 0.003 & 0.374 & 0.960 & 0.923(0.002) & 0.004 & 0.370 & 0.961 \\
$M_{i}$         & 0.929(0.002) & 0.005 & 0.358 & 0.964 & 0.928(0.002) & 0.003 & 0.361 & 0.963 & 0.930(0.002) & 0.003 & 0.357 & 0.964 \\
$M_{z}$         & 0.933(0.002) & 0.005 & 0.353 & 0.966 & 0.932(0.002) & 0.003 & 0.357 & 0.965 & 0.933(0.002) & 0.003 & 0.353 & 0.966 \\
SFR             & 0.697(0.004) & 0.039 & 0.274 & 0.835 & 0.694(0.003) & 0.038 & 0.276 & 0.833 & 0.695(0.004) & 0.039 & 0.274 & 0.834 \\
$Z_{\rm{gas}}$  & 0.771(0.002) & 0.023 & 0.136 & 0.878 & 0.769(0.002) & 0.028 & 0.137 & 0.877 & 0.772(0.002) & 0.025 & 0.136 & 0.879 \\
\hline
\noalign{\vskip 2pt}
\multicolumn{5}{c}{} & \multicolumn{4}{c}{\textit{Part 2 of 2}} & \multicolumn{4}{c}{} \\
\cline{2-5}\cline{6-9}\cline{10-13}
 & \multicolumn{4}{c}{Simulation Particle} & \multicolumn{4}{c}{Subhalo Anisotropy} & \multicolumn{4}{c}{Maximal} \\
Output & $R^2$ & $\Delta R^2$ & MAE & $\rho$ & $R^2$ & $\Delta R^2$ & MAE & $\rho$ & $R^2$ & $\Delta R^2$ & MAE & $\rho$ \\
\hline
$M_{\rm{star}}$ & 0.956(0.001) & 0.006 & 0.122 & 0.978 & 0.958(0.001) & 0.006 & 0.120 & 0.979 & 0.959(0.001) & 0.006 & 0.119 & 0.979 \\
$M_{\rm{gas}}$  & 0.886(0.003) & 0.005 & 0.130 & 0.942 & 0.896(0.003) & 0.018 & 0.127 & 0.947 & 0.895(0.003) & 0.009 & 0.126 & 0.946 \\
$M_{U}$         & 0.847(0.003) & 0.010 & 0.513 & 0.920 & 0.848(0.003) & 0.007 & 0.513 & 0.921 & 0.848(0.003) & 0.007 & 0.512 & 0.921 \\
$M_{B}$         & 0.891(0.002) & 0.008 & 0.431 & 0.944 & 0.892(0.003) & 0.005 & 0.428 & 0.945 & 0.893(0.002) & 0.006 & 0.427 & 0.945 \\
$M_{V}$         & 0.916(0.002) & 0.006 & 0.382 & 0.957 & 0.918(0.002) & 0.004 & 0.379 & 0.958 & 0.918(0.002) & 0.005 & 0.378 & 0.958 \\
$M_{K}$         & 0.938(0.001) & 0.005 & 0.359 & 0.968 & 0.939(0.001) & 0.004 & 0.355 & 0.969 & 0.940(0.001) & 0.004 & 0.353 & 0.969 \\
$M_{g}$         & 0.901(0.002) & 0.007 & 0.412 & 0.949 & 0.902(0.002) & 0.005 & 0.409 & 0.950 & 0.903(0.002) & 0.005 & 0.408 & 0.950 \\
$M_{r}$         & 0.923(0.002) & 0.006 & 0.369 & 0.961 & 0.925(0.002) & 0.004 & 0.365 & 0.962 & 0.925(0.002) & 0.005 & 0.364 & 0.962 \\
$M_{i}$         & 0.930(0.002) & 0.005 & 0.356 & 0.964 & 0.932(0.002) & 0.004 & 0.352 & 0.965 & 0.932(0.002) & 0.004 & 0.350 & 0.966 \\
$M_{z}$         & 0.934(0.002) & 0.005 & 0.351 & 0.966 & 0.935(0.002) & 0.004 & 0.347 & 0.967 & 0.936(0.001) & 0.004 & 0.346 & 0.967 \\
SFR             & 0.699(0.004) & 0.039 & 0.273 & 0.836 & 0.701(0.004) & 0.048 & 0.272 & 0.837 & 0.701(0.004) & 0.038 & 0.272 & 0.837 \\
$Z_{\rm{gas}}$  & 0.781(0.002) & 0.023 & 0.133 & 0.884 & 0.781(0.001) & 0.029 & 0.134 & 0.884 & 0.792(0.001) & 0.026 & 0.130 & 0.890 \\
\enddata
\tablecomments{Each column denotes the CV $R^2$ mean (standard deviation), $\Delta R^2$ (train$-$test), CV MAE mean, and CV $\rho$ mean.}
\end{deluxetable*}

\section{Softening Length and Subhalo Properties}
\label{AppC}
\begin{figure*}[t]
\centering
\includegraphics[width=0.8\textwidth]{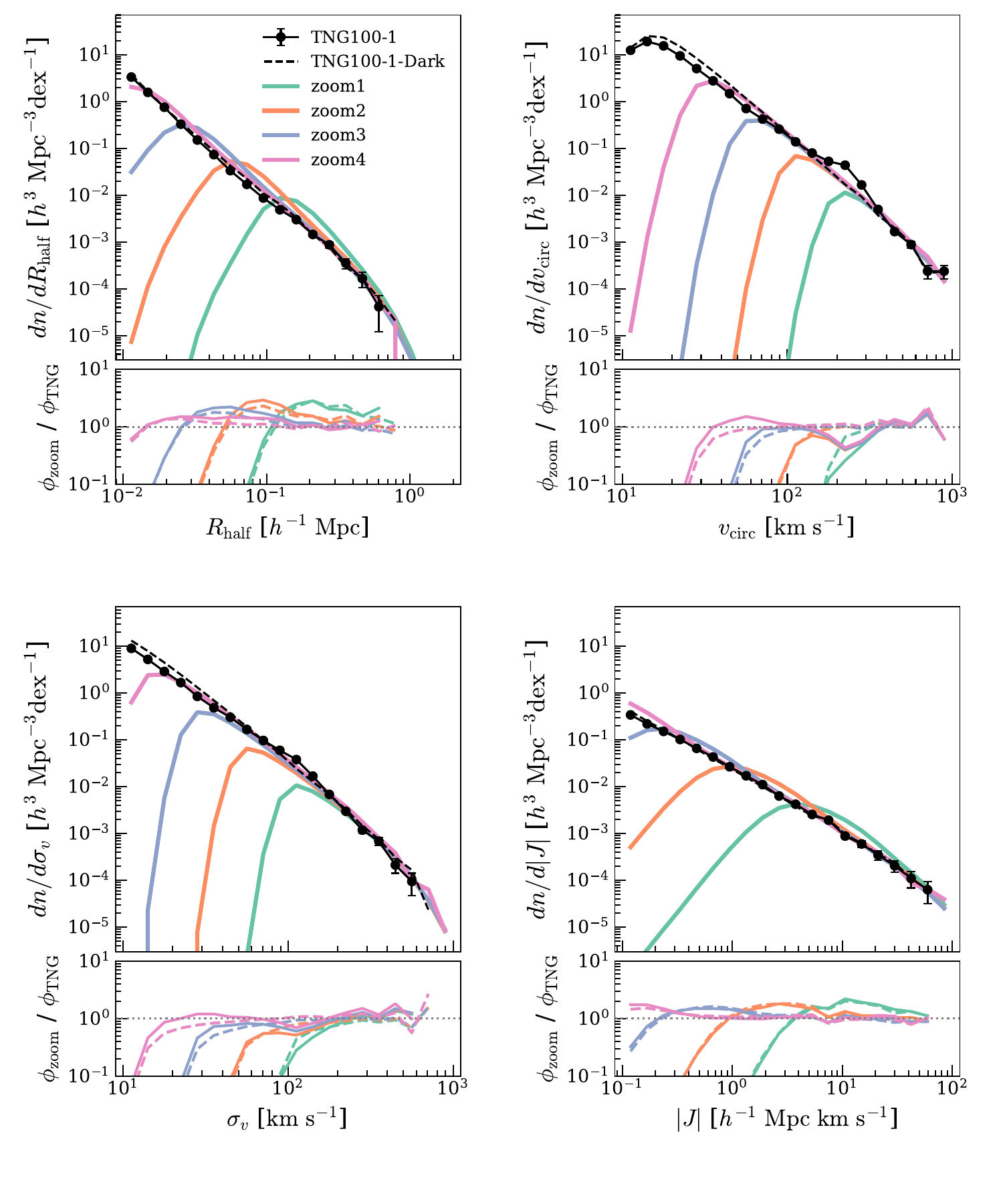}
\caption{Distribution functions of $R_{\rm{half}}$, $v_{\rm{circ}}$, $\sigma_v$, and $|J|$ in zoomed-in simulations (colored lines) and TNG100-1 (black solid line). TNG100-1-Dark (black dashed line), the DM-only counterpart simulation of TNG100-1, is shown together to show the difference between hydro and DM-only simulations. $R_{\rm{half}}$ and $|J|$ show systematic offsets (factor of 2), but $v_{\rm{circ}}$ and $\sigma_v$ show less systematic deviation. However, bumps are present around $v_{\rm{circ}}\approx200\,\rm{km}\,\rm{s}^{-1}$ and $\sigma_v\approx100\,\rm{km}\, \rm{s}^{-1}$, which are also observed in TNG100-1-Dark.  
\label{fig:AppendixB_SubhaloSize}}
\end{figure*}

In this appendix, we compare the physical properties of subhalos other than mass across different simulations. We particularly focus on half-mass radius $R_{\rm{half}}$, maximum circular velocity $v_{\rm{circ}}$, velocity dispersion $\sigma_v$, and spin $|J|$ that are related to the choice of softening length. In Figure \ref{fig:AppendixB_SubhaloSize}, the distribution function of each property is shown along with those calculated from the TNG100-1 and TNG100-1-Dark simulations. The subhalo size increases as the resolution worsens, and the spin is also systematically larger accordingly. However, $v_{\rm{circ}}$ and $\sigma_v$ show a smaller difference compared to the radius and spin.

One noticeable difference other than the resolution effect is the kink observed at $v_{\rm{circ}}\approx200\,\rm{km}\,\rm{s}^{-1}$ and $\sigma_v\approx100\,\rm{km}\,\rm{s}^{-1}$. It stems from the baryonic physics because the DM counterpart simulation (TNG100-1-Dark) does not exhibit such a kink in its distributions. As $v_{\rm{circ}}$ and $\sigma_v$ play a crucial role in predicting baryonic properties, exploring and correcting for the hydrodynamical effect is essential for painting DM subhalos with ML.

\begin{figure*}[t]
\centering
\includegraphics[width=0.8\textwidth]{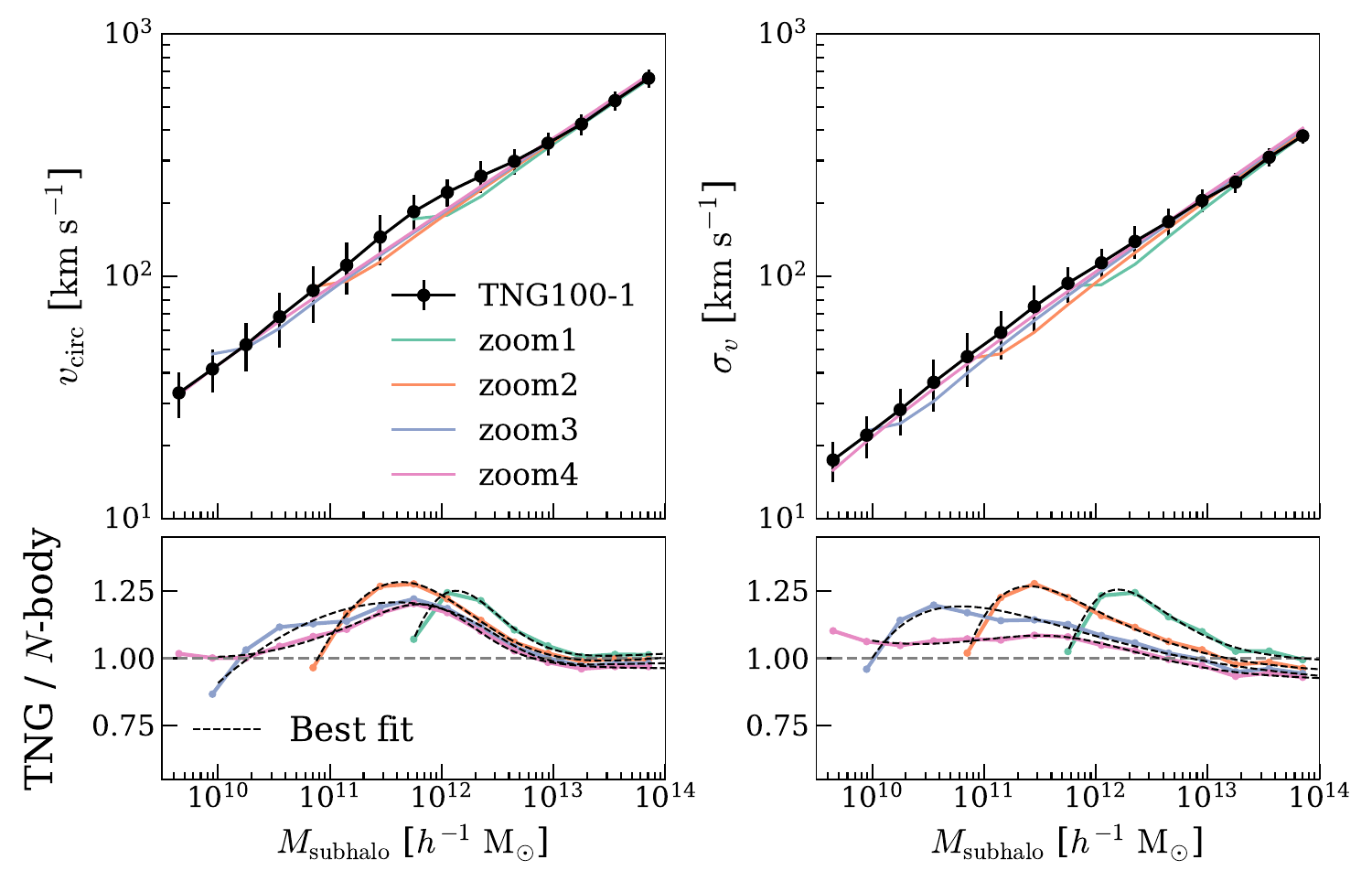}
\caption{Mass-dependent distribution of $v_{\rm{circ}}$ and $\sigma_v$. The black line is the measurement from TNG100-1, and the colored lines show the measurements from zoomed-in simulations. The error bars show the dispersion in each bin. We fit the ratio between $N$-body simulations and TNG as a function of mass and use it to scale when applying MSSM to the $N$-body simulations. Best-fit functions are shown as black dashed lines.
\label{fig:AppendixB_Vmax_Sigma_scaling}}
\end{figure*}

To match the distribution of $v_{\rm{circ}}$ and $\sigma_v$, we use a scaling factor as a function of mass. While the dependence is not purely on mass, this simple mass-dependent scaling approximately captures the trend driven by the interplay between stellar/AGN feedback and the depth of the halo potential \citep[see e.g.,][]{DiCintio2014, Schaller2015, Anbajagane2022, Sorini2025}. We calculate the mean and dispersion of those parameters for each mass bin equally spaced in $\log(M_{\rm{subhalo}}\,[h^{-1}\,M_{\odot}])=[9.5, 14]$. The result is shown as the black line (TNG100-1) and colored lines ($N$-body) in Figure \ref{fig:AppendixB_Vmax_Sigma_scaling}. We fit the ratio between those two (bottom panel) and apply it before the baryonic properties are predicted (Section \ref{sec:Sec4.3}). The universal fitting function $f(m)$ given below is an empirical choice.
\begin{equation}
    f_{N\rm{-body} \rightarrow \rm{TNG100-1}}(m)=A + \dfrac{am^2+bm+c}{1+\exp\left[\frac{m-\mu}{\sigma}\right]}\,,
\end{equation}
where $m\equiv\log(M_{\rm{subhalo}}\,[h^{-1}\,M_{\odot}])$, and the function approaches a constant at the high-mass end.\\

\section{Full Result of MSSM Application to the $N$-body Simulation}
\label{AppD}
\begin{figure*}[t]
\centering
\includegraphics[width=0.7\textwidth]{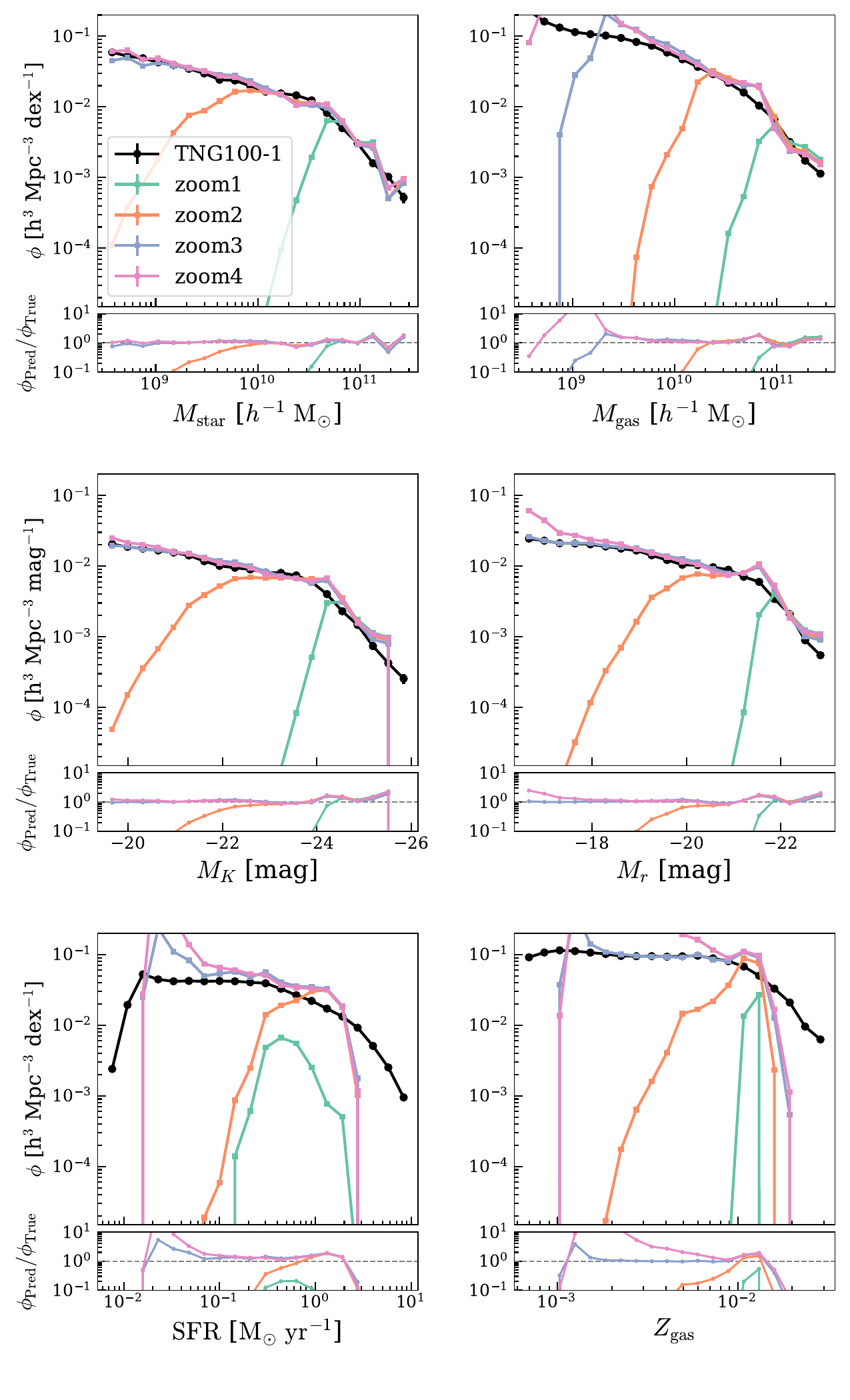}
\caption{Distribution functions of predicted baryonic properties from the $N$-body simulations.
\label{fig:AppendixD_NASIM_AllBaryon}}
\end{figure*}

In this appendix, we show the distribution functions of predicted baryonic properties from the $N$-body simulations. The results are shown in Figure \ref{fig:AppendixD_NASIM_AllBaryon}.\\

\section{Impact of Random Scatter on the Redshift Distribution of Mock Galaxies}
\label{AppE}
\begin{figure}[t]
\centering
\includegraphics[width=0.99\linewidth]{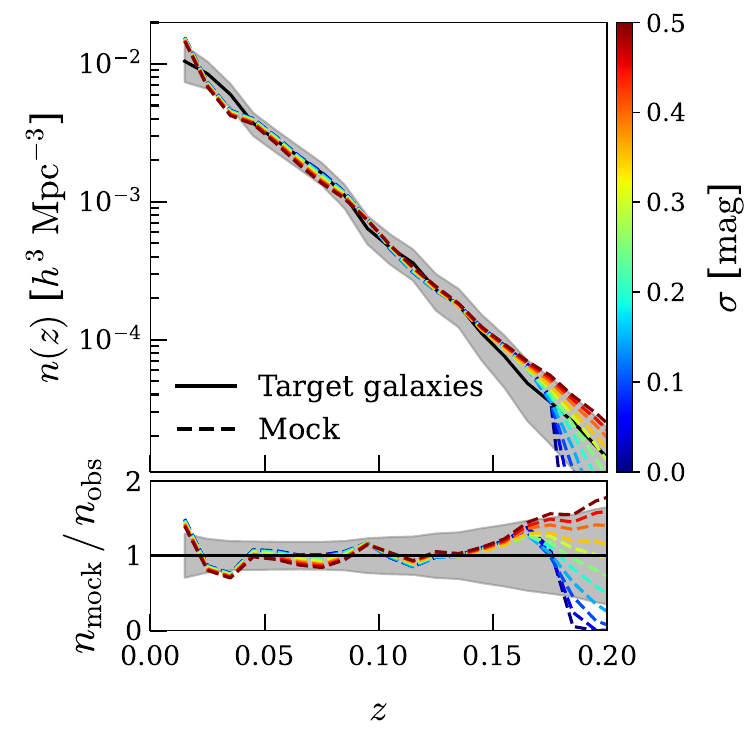}
\caption{The redshift distribution of galaxies with $K_S\leqslant13.75$ from the target galaxy catalog (solid) and the mock data (dashed). The color bar shows the standard deviation of the random Gaussian scatter added to the mock $M_{K_S}$.
\label{fig:AppendixE}}
\end{figure}

In this appendix, we demonstrate the redshift distribution of mock galaxies, depending on the random scatter added to the predicted $M_{K_S}$ (Section \ref{sec:Sec4.4}). Figure \ref{fig:AppendixE} shows the redshift distribution of galaxies with $K_S\leqslant13.75$. The black line denotes the measurement with the target galaxy catalog (identical to Figure \ref{fig:Fig3_z_nz}), and the colored lines show the measurements from the mock galaxy catalog. The colors show the standard deviation of the Gaussian scatter added to the predicted $M_{K_S}$. The mock data agree with the observation with $\sigma\gtrsim0.3\,\textrm{mag}$.

\bibliography{sample701}{}
\bibliographystyle{aasjournalv7}



\end{document}